\documentclass[final,onefignum,onetabnum]{siamart220329}

\usepackage{adjustbox}
\usepackage[linesnumbered,ruled,vlined,algo2e]{algorithm2e}
\usepackage{amsfonts}
\usepackage{bm}
\usepackage[T1]{fontenc}
\usepackage{lmodern}
\usepackage{tikz}
\usetikzlibrary{decorations.markings}

\newcommand{\edgelong}[3]{\psi(#1,#2,#3)}
\newcommand{\edge}[3]{\psi_{#1#2}(#3)}

\renewcommand{\vec}[1]{\mathbf{#1}}
\newcommand{\R}{\mathbb{R}}

\newcommand{\bx}{{\bm x}}

\newcommand{\ieeeBits}[1]{\ensuremath{\textsc{ie}^3({#1})}}

\newcommand{\bitsLowerMantissa}{\ieeeBits{[0{-}25]}}
\newcommand{\bitsUpperMantissa}{\ieeeBits{[26{-}51]}}
\newcommand{\bitsExponent}{\ieeeBits{[52{-}62]}}
\newcommand{\bitsSign}{\ieeeBits{63}}

\graphicspath{{img/poisson2d}{img/mgglap}}

\headers{Modifying Asynchronous Jacobi for Data Corruption}{C. J. Vogl, Z. Atkins, A. Fox, A. Mi\k{e}dlar, C. Ponce}

\title{Modifying the Asynchronous Jacobi Method for Data Corruption Resilience.
	\thanks{
		\funding{This work was performed under the auspices of the U.S. Department of Energy by Lawrence Livermore National Laboratory under Contract DE-AC52-07NA27344 and was supported by the LLNL-LDRD Program under Project No. 21-FS-007 and 22-ERD-045, LLNL-JRNL-853510-DRAFT.
	  The work of Agnieszka Mi\k{e}dlar was supported by the NSF grants DMS-2144181 and DMS-2324958.}
	}
}

\author{Christopher J. Vogl\thanks{Lawrence Livermore National Laboratory,
		Livermore, CA (\email{vogl2@llnl.gov})}
	\and Zachary Atkins\thanks{University of Colorado, Boulder, CO (\email{zach.atkins@colorado.edu})}
	\and Alyson Fox \thanks{Lawrence Livermore National Laboratory,
		Livermore, CA (\email{fox33@llnl.gov})}
	\and Agnieszka Mi\k{e}dlar\thanks{Virginia Tech, Blacksburg, VA (\email{amiedlar@vt.edu})}
	\and Colin Ponce \thanks{Lawrence Livermore National Laboratory,
		Livermore, CA (\email{ponce11@llnl.gov})}}

\begin{document}
\maketitle

\begin{abstract}
	Moving scientific computation from high-performance computing (HPC) and cloud computing (CC) environments to devices on the edge, i.e., physically near instruments of interest, has received tremendous interest in recent years.
	Such edge computing environments can operate on data in-situ, offering enticing benefits over data aggregation to HPC and CC facilities that include avoiding costs of transmission, increased data privacy, and real-time data analysis.
	Because of the inherent unreliability of edge computing environments, new fault tolerant approaches must be developed before the benefits of edge computing can be realized.
	Motivated by algorithm-based fault tolerance, a variant of the asynchronous Jacobi (ASJ) method is developed that achieves resilience to data corruption by rejecting solution approximations from neighbor devices according to a bound derived from convergence theory.
	Numerical results on a two-dimensional Poisson problem show the new rejection criterion, along with a novel approximation to the shortest path length on which the criterion depends, restores convergence for the ASJ variant in the presence of certain types data corruption.
	Numerical results are obtained for when the singular values in the analytic bound are approximated.
	A linear system with a more dense sparsity pattern is also explored.
	All results indicate that successful resilience to data corruption depends on whether the bound tightens fast enough to reject corrupted data before the iteration evolution deviates significantly from that predicted by the convergence theory defining the bound.
	This observation generalizes to future work on algorithm-based fault tolerance for other asynchronous algorithms, including upcoming approaches that leverage Krylov subspaces.
\end{abstract}

\section{Introduction}
Recent years have seen a proliferation of \textit{edge devices}, i.e., streamlined computing devices that provide an entry point to the individual instruments in their vicinity.
Modern infrastructure includes a wide range of such devices, from smart residential thermostats to industrial smart grid meters.
These devices, along with wearable healthcare devices and content delivery systems, are motivating a push of computation beyond the walls of high-performance (HPC) and cloud computing (CC) facilities onto the edge devices themselves.
Consider, as an example, the benefits of enabling smart power grid devices to operate autonomously when the central operator is disabled due to a natural disaster or cyber-physical attack.
The capability provided by edge computing environments to operate without a single point of failure or on data in-situ is appealing to real-time system operators.
Unfortunately, the benefits of edge computing cannot be realized before the inherent unreliability of edge devices is addressed.
Modern scientific computing algorithms typically assume that data will not be corrupted as the algorithm is executed.
HPC and CC platforms provide such data integrity by utilizing fault management techniques.
Checkpointing and redundant computation are cornerstones of fault management techniques in HPC and CC, and are an integral part of n-modular redundancy \cite{canal}, n-version programming \cite{mukwevho}, majority voting \cite{canal}, and redundant cloud servers \cite{mukwevho} techniques.
The frequency of checkpointing is typically chosen to avoid restarting from a checkpoint created long before the fault occurs while keeping the cost of synchronization and storage reasonable.
Similarly, the amount of redundancy is typically chosen to avoid having all redundant entities experience a fault at the same time while keeping the cost of storage and flops reasonable.
Thus, checkpointing and redundancy are not practical for edge computing environments where synchronization is typically expensive, and data storage and/or flops are limited.

One promising alternative fault management strategy is the class of \textit{algorithm based fault tolerant} (ABFT) methods.
The general idea is to leverage the structure or expected behavior of the algorithm to detect, mitigate, and/or recover from faults such as data corruption.
Examples of ABFT schemes include methods for the fast Fourier transform \cite{liang_correcting_2017}, matrix multiplication \cite{wu_fault_2011}, Krylov-based iterative methods \cite{chen_online_2013}, and the synchronous Jacobi method \cite{anzt_2015}.
Focusing on the iterative methods, the work presented in \cite{chen_online_2013} uses the orthogonality of projections onto Krylov spaces for detection of faults,
while \cite{anzt_2015} utilizes the contraction mapping property of stationary iterative methods.
Unfortunately, those ABFT approaches are for iterative methods that require frequent synchronizations, making them impractical for edge computing environments due to network latency, heterogeneous nodes, and nonpersistent nodes/links.
Asynchronous methods remove the need for global synchronization after each iteration, allowing them to outperform their synchronous counterparts in high-latency environments (see, e.g., \cite{wolfson_jacobi}).
This makes asynchronous methods very appealing for edge computing; however, the authors are aware of only two existing asynchronous methods with ABFT strategies: the robust alternating direction method of multipliers (ADMM) \cite{robust_admm} and the robust push-sum algorithm \cite{olshevsky}.

Solving the nonlinear systems that are being pushed to edge computing environments, such as distribution system state estimation \cite{wang_distribution_2019} and machine learning for autonomous vehicles \cite{liu_edge_2019} and smart agriculture \cite{kalyani_systematic_2021}, typically require the solution of one or more linear systems (e.g., direction vectors in the Newton-Rhapson method).
To address the resulting need for an asynchronous linear solver with ABFT for such tasks, we modify the asynchronous Jacobi (ASJ) method \cite{asynchronous,hook_performance_2018,wolfson_jacobi,anzt_2015,chaotic} with a rejection criterion based on the convergence properties of ASJ.
In \cite{robust_admm}, such a rejection criterion is developed where data from neighboring nodes is rejected if the difference between successive data obtained from a neighbor exceeds a bound derived from the convergence theory for ADMM.
Here, a similar criterion is used but with a novel bound derived for the ASJ method based on the ASJ convergence theory developed in \cite{hook_performance_2018}.
Because the bound for ASJ depends on time-dependent global information, specifically the shortest path length, a novel local approximation to that global information is also developed.
It is worth noting that the choice of the ASJ method for modification is motivated by (i) the authors are not aware of another asynchronous linear solver with established convergence theory, (ii) while the Jacobi method is known to scale poorly to large and ill-conditioned systems, ASJ can be sufficient for small problems that appear in edge environments (e.g., state estimation in a neighborhood), and (iii) the observations and understanding of the resilience modification for ASJ can be generalized to upcoming asynchronous Krylov-based solvers, such as the one recently developed in \cite{erlandson_resilient_2023}.

This paper is organized as follows.
Section \ref{sec:problemformulation} formulates the problem, introduces the notation and important definitions, and discusses the nature of data corruption to be investigated.
Section \ref{sec:modification} develops our resilience enabling technique.
Section \ref{sec:results} presents numerical results verifying the implementation of the method and demonstrating the effectiveness of the proposed rejection technique in the presence of various forms of data corruption.
An empirical sensitivity study of the rejection criterion parameters is also presented, with a discussion on potential improvements to the local approximation of global information used in the criterion.
The evaluation of the method is extended to a second linear system with a more dense sparsity pattern.
Section \ref{sec:conclusions} summarizes the outcomes of the paper and discusses how the results might generalize to ongoing and future work.

\section{Problem Statement}
\label{sec:problemformulation}
Solutions of linear systems are ubiquitous in modern scientific computing algorithms, defining search directions in both iterative and nonlinear solvers.
Thus, consider solving the linear system
\begin{gather}
  A \vec{x} = \vec{b},
  \label{eq:axb}
\end{gather}
for $\vec{x} \in \R^m$, where $A \in \R^{m \times m}$ and $\vec{b} \in \R^m$.
Assume that $A$ is non-singular so that a unique solution to \eqref{eq:axb} exists.
The asynchronous Jacobi method is an iterative solver for \eqref{eq:axb} in that successive approximations to the solution $\vec{x}$ are formed across $N$ computational nodes.
Denote by $\vec{x}_i \in \R^{m_i}$, $m_i \leq m$, the partition of $\vec{x}$ that node $i$ is approximating.
Let $I \in \R^{m \times m}$ and $D \in \R^{m \times m}$ be the identity matrix and the diagonal matrix containing the diagonal elements of $A$, respectively.
The update equation that defines the successive approximations computed by node $i$, denoted $\vec{x}_i^0$, $\vec{x}_i^1$, etc., can now be expressed as
\begin{gather}
  \vec{x}_i^\kappa = \sum_{j=1}^N M_{ij} \vec{x}_j^{\edgelong{i}{j}{\kappa}} + \vec{c}_i, \quad \kappa = 1,2,\ldots
  \label{eq:asj}
\end{gather}
where $M_{ij} \in \R^{m_i \times m_j}$ is the partition of the Jacobi iteration matrix $M := I-D^{-1}A$ with rows that correspond to $\vec{x}_i$ and columns that correspond to $\vec{x}_j$, and $\vec{c}_i \in \R^{m_i}$ is the partition of $\vec{c} := D^{-1} \vec{b}$ with elements that correspond to $\vec{x}_i$.
The index function $\psi$ is defined by $\edgelong{i}{j}{\kappa} = \lambda$ if node $i$ uses node $j$’s $\lambda$-th approximation in the computation of its $\kappa$-th approximation.
All matrix, vector, and matrix-vector operations, including the communication of $\vec{x}_i^\kappa$ and $\vec{x}_j^{\edgelong{i}{j}{\kappa}}$, are to be considered block operations.

The general form of \eqref{eq:asj} defines a class of chaotic or asynchronous iterative methods, first introduced by Chazan and Miranker \cite{chaotic}, that generalize classic relaxation methods to allow each compute node to perform a new iteration immediately after the previous iteration has completed.
Chazan and Miranker provide the sufficient condition to guarantee convergence of any relaxation scheme of the form \eqref{eq:asj}: that the spectral radius of the absolute value of the global iteration matrix, $M$, is bounded below one, i.e., $\rho(|M|)<1$, where $|M|$ is defined by taking the absolute value of each element in the matrix.
However, the authors in \cite{chaotic} assume that the values of $\vec{x}_j^{\edgelong{i}{j}{\kappa}}$ sent by node $j$ are the same as those received by node $i$.
Such assumption can become invalid in emerging computing environments that do not provide the guarantees of current high performance computing systems.
Thus, the goal of this work is to modify \eqref{eq:asj} to ensure, or at least encourage, convergence even if data corruption results in either (i) the values of $\vec{x}_j^{\edgelong{i}{j}{\kappa}}$ received by node $i$ being different than those sent by node $j$ or (ii) the values of $\vec{x}_j^\kappa$ stored on node $j$ being altered.
As a convenience to the reader, Table \ref{table:notation} summarizes the notation used herein, as well as the location where the notation is first mentioned.

\begin{table}[htbp]
	\caption{Notation Table}
	\begin{center}
		\begin{adjustbox}{width=1\textwidth}
			\begin{tabular}{|c l c|}
				\hline
				Symbol & Description & Location \\
				\hline
				\hline
				$N$ & number of computational nodes &  Section \ref{sec:problemformulation} \\
				$A$ & square system matrix ($\mathbb{R}^{m \times m}$) with real components &  Section \ref{sec:problemformulation} \\
				$\vec{x}$ , $\vec{b}$  &  vectors of real components ($\mathbb{R}^m$) & Section \ref{sec:problemformulation} \\
				$\vec{x}_i $ & vector ($\R^{m_i}$) containing the $i$-th partition of $\vec{x}$ assigned to the $i$-th computational node & Section \ref{sec:problemformulation} \\
				$D$ & diagonal matrix ($\R^{m \times m}$) whose elements are the diagonal entries of $A$ &  Section \ref{sec:problemformulation} \\
				$M$ & Jacobi iteration matrix $D^{-1}A$ ($\mathbb{R}^{m \times m}$) &  Section \ref{sec:problemformulation} \\
				$M_{i,j}$ & partition of $M$ ($\mathbb{R}^{m_i \times m_j}$) with rows corresponding to $\vec{x}_i$ and columns corresponding to $\vec{x}_j$ &  Section \ref{sec:problemformulation} \\
				i,j,k  & indexing of partitions &  Section \ref{sec:problemformulation} \\
				$\lambda, \kappa$  & indexing of iterations &  Section \ref{sec:problemformulation}\\
				$\vec{x}_{i}^\kappa $ & the $\kappa$-th iteration of the approximation to vector $\vec{x}$ on node $i$ ($\R^{m_i}$) & Section \ref{sec:problemformulation}\\
				$p$ & the probability of a bit flip in a communicated element  & Section~\ref{sec:natural-corruption} \\
				\(\omega_f\),  \(\omega_r\)  & the time to failure and recovery time &  Section~\ref{sec:malevolent-corruption} \\
				$\delta$ & an offset that is sampled from a Gaussian distribution with a positive mean & Section~\ref{sec:malevolent-corruption} \\
				$\nu_i(t)$ & iteration index such that $\vec{x}_i^{\nu_i(t)}$ is the most recent approximation of $\vec{x}_i$ at time $t$ & Section \ref{sec:modification} \\
				$\widetilde{\vec{x}}(t)  $ & global approximate solution ($\R^m$) at time $t$ such that $\vec{\widetilde{x}}_i(t) = \vec{x}_i^{\nu_i(t)}$, $i=1,\ldots,N$ & Section \ref{sec:modification}\\
				$\vec{x}^*$ & global exact solution ($\R^m$) to $A\vec{x} = \vec{b}$ &  Section \ref{sec:modification}\\
				$\vec{e}(t)$ & global error ($\R^m$) at time $t$ such that $\vec{e}_i(t) =\widetilde{\vec{x}}_i(t) - \vec{x}_i^*$ &  Section  \ref{sec:modification}\\
				$\Omega(t):= \Omega(\psi,M,t)$ & error operator ($\R^{m \times m}$) such that $\vec{e}(t) = \Omega(t)\vec{e}(0)$ & Section \ref{sec:modification}\\
				$\mathcal{G}(\mathcal{V}, \mathcal{E} )$ & directed acyclic graph with nodes $\mathcal{V}$ and edges $\mathcal{E}$ &  Section \ref{sec:modification} \\
				$s(t)$, $l(t)$ & shortest and longest paths in $\mathcal{G}$, respectively & Section \ref{sec:modification} \\
				$\edge{i}{j}{\kappa}:=\edgelong{i}{j}{\kappa}$ & iteration index of the update from node $j$ which node $i$ uses to compute its $\kappa$-th update &  Section \ref{sec:modification} \\
				$\zeta_{ij}(t)$ & $\arg\max_\kappa {\psi_{ij}(\kappa ) < \psi_{ij}(\nu_i(t))}$ & Section \ref{sec:modification} \\
				$\tau_{ij}[\kappa]$ & time at which the solution approximation existed on node $j$ that will later be used to form $\vec{x}_i^{\kappa}$ & Section \ref{sec:modification} \\
				$\sigma_{min}(A),\sigma_{max}(A)$ &  smallest and largest singular value of $A$, respectively & Section \ref{sec:modification} \\
				$\tilde{s}_i(t)$ & approximate shortest path  & Section \ref{sec:modification} \\
				$d\vec{x}_i^{\nu_i(t)} := \vec{x}_i^{\nu_i(t)} - \vec{x}_i^{\nu_i(t)-1}$  & difference ($\R^{m_i}$) between two successive solution approximations on node $i$ & Section \ref{sec:modification} \\
				$\epsilon$ & a user defined tolerance for the stopping criteria & Section  \ref{sec:modification} \\
				\hline
			\end{tabular}
		\end{adjustbox}
	\end{center}

	\label{table:notation}
\end{table}

\subsection{Natural Data Corruption}
\label{sec:natural-corruption}

  The first data corruption model is motivated by bit flips occurring in network hardware memory that alter data as it is in transit.
  This natural data corruption is modeled as a random process where each component of transmitted data is affected by a bit flip with a fixed probability $p \in (0,1)$.
  The bit flips themselves are performed either on \textsc{ieee} 754 double precision (64 bit) floating point \cite{IEEE754_spec} or on 32 bit signed integer numbers.
  The affected bit index is sampled from various uniform integer distributions, then the bit flip is performed directly.
  In extremely rare cases, this method of performing bit flips on double precision numbers can result in the special floating point values \texttt{NaN} or \texttt{inf}.
  It is worth noting that this data corruption approach mirrors the model of Anzt et al. \cite{anzt_2015}, where a fixed number of bit flips are introduced per iteration to the entries of the iteration matrix \(M\) during the matrix-vector product in each iteration, which may corrupt up to \(1\%\) of updates to the elements of the solution vector.
  Instead, we choose to corrupt the elements of the transmitted solution vector directly at a fixed probability \(p\in(0,1)\), i.e., corruption is applied with probability $p$ to each transmitted data element.

\subsection{Malevolent Data Corruption}
\label{sec:malevolent-corruption}

  The second data corruption model is motivated by intentional corruption caused by a malicious actor who has gained intermittent access to a device to manipulate the result of a calculation.
  This malevolent data corruption is modeled as a periodic process where each agent is considered to be in either a ``normal'' or a ``degraded'' state.
  When in a ``normal'' state, the new approximate solution is unaltered.
  After \(\omega_f\) seconds have passed, the agent is compromised and enters a ``degraded'' state.
  While in the ``degraded'' state, the impacted data on an agent is corrupted by adding an offset to all solution elements.
  Note that such non-transient corruption, i.e., overwriting of the local solution data, presents a more challenging recovery scenario than transient corruption.
  This offset is sampled from a Gaussian distribution with a positive mean \(\delta\) and a standard deviation of \(0.5\delta\).
  The repeated application of these offsets will gradually increase the magnitude of the corrupted elements of the solution vector, absent any mitigation strategy.
  We choose the standard deviation \(0.5\delta\) to ensure that \(95\%\) of the sampled offsets will be greater than zero regardless of the choice of \(\delta\).
  After \(\omega_r\) seconds have passed, the agent is secured and returns to a ``normal'' state.

\section{Corruption Resilience Modification}
\label{sec:modification}
To improve data corruption resilience in the asynchronous Jacobi method \eqref{eq:asj}, we take the approach of inspecting incoming data before it is used to form the next approximation $\vec{x}_i^\kappa$ in \eqref{eq:asj}.
If the data is identified as corrupted, it is rejected by being excluded from contributing to the next solution approximation.
We will use the convergence theory established by Hook and Dingle \cite{hook_performance_2018} to derive our rejection criterion.
The authors in \cite{hook_performance_2018} derive an error bound using metrics of the evolution of the solution approximations computed by each node and the communication pattern between nodes.
They accomplish this task by casting the algorithm evolution as a directed acyclic graph, whose vertices are the solution approximations computed by each node and edges indicate when a solution approximation is used to compute a latter solution approximation.
Using a similar notation, we will summarize the components used to form the rejection criterion below.

\begin{figure}[htbp]
	\centering
	\begin{tikzpicture}
	
\newcommand\Xpos{2}
\newcommand\Ypos{0}
\newcommand\Tone{5.5}

	\draw[red,fill=red] (0,\Ypos) circle [fill,radius=1.5mm];
	\draw (0,\Ypos)  node[below=2ex] {$\vec{x}_1^0$};
	\draw[red,fill=red] (2,\Ypos) circle [fill,radius=1.5mm];
	\draw[red,fill=red] (4,\Ypos) circle [fill,radius=1.5mm];
	\draw[red,fill=red] (6.5,\Ypos) circle [fill,radius=1.5mm];

	\draw[red,fill=red] (0,\Xpos) circle [fill,radius=1.5mm];
	\draw (0,\Xpos)  node[above=2ex] {$\vec{x}_2^0$};
	\draw[red,fill=red] (3,\Xpos) circle [fill,radius=1.5mm];
	\draw[red,fill=red] (4.5,\Xpos) circle [fill,radius=1.5mm];
	\draw[red,fill=red] (8,\Xpos) circle [fill,radius=1.5mm];
	
\begin{scope}[very thick,decoration={
		markings,
		mark=at position 0.5 with {\arrow{>}}}
	] 
	\draw[postaction={decorate}] (0,\Ypos)--(2,\Ypos)  node[below=2ex] {$\vec{x}_1^1$};
	\draw[postaction={decorate}] (2,\Ypos)--(4,\Ypos)  node[below=2ex] {$\vec{x}_1^2$};
	\draw[postaction={decorate}] (4,\Ypos)--(6.5,\Ypos)  node[below=2ex] {$\vec{x}_1^3$};

	\draw[postaction={decorate}] (0,\Xpos)--(3,\Xpos)  node[above=2ex] {$\vec{x}_2^1$};
	\draw[postaction={decorate}] (3,\Xpos)--(4.5,\Xpos)  node[above=2ex] {$\vec{x}_2^2$};
	\draw[postaction={decorate}] (4.5,\Xpos)--(8,\Xpos)  node[above=2ex] {$\vec{x}_2^3$};

	\draw[postaction={decorate}] (0,\Xpos)--(2,\Ypos) ;
	\draw[postaction={decorate}] (3,\Xpos)--(4,\Ypos) ;
	\draw[postaction={decorate}] (4.5,\Xpos)--(6.5,\Ypos) ;

	\draw[postaction={decorate}] (2,\Ypos)--(3,\Xpos) ;
	\draw[postaction={decorate}] (2,\Ypos)--(4.5,\Xpos) ;
	\draw[postaction={decorate}] (4,\Ypos)--(8,\Xpos) ;
\end{scope}

\draw [very thick,-> ] (6.5,\Xpos) -- (9.5,\Xpos);
\draw [very thick,->] (6,\Ypos) -- (9.5,\Ypos);

\draw [dashed]  (\Tone,\Xpos+1 )-- (\Tone,\Ypos-1) node[below=2ex] {$t = t_1$};

\end{tikzpicture}
	\caption{Directed acyclic graph $\mathcal{G}(\mathcal{V}, \mathcal{E})$ illustrating an example two-node evolution of the solution approximations $\vec{x}_1^{\nu_1(t)}$ and $\vec{x}_2^{\nu_2(t)}$.}
	\label{fig:dag}
\end{figure}
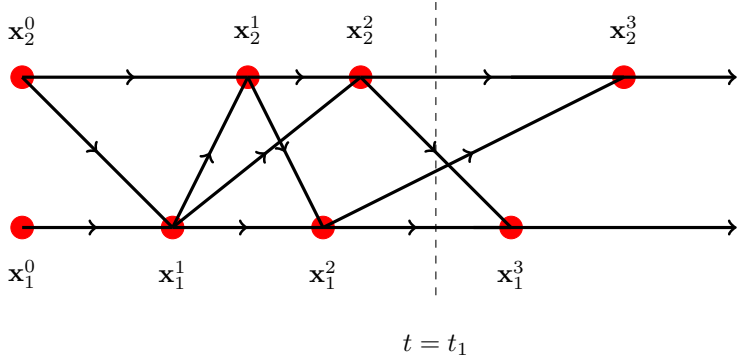

Define $\nu_i(t)$ so that $\vec{x}_i^{\nu_i(t)}$ is the solution approximation on node $i$ at time $t$.
A global solution approximation at time $t$, denoted by $\vec{\widetilde{x}}(t)$, is defined block-wise as $\vec{\widetilde{x}}_i(t) = \vec{x}_i^{\nu_i(t)}$, $i=1,\ldots,N$.
With $\vec{x}^*$ being the exact solution of \eqref{eq:axb}, the global error at time $t$ is defined as $\vec{e}(t) = \vec{\widetilde{x}}(t) - \vec{x}^*$.
Denote the error operator $\Omega(\psi,M,t)$ such that $\vec{e}(t) = \Omega(\psi,M,t)\vec{e}(0)$.
The properties of $\Omega(\psi,M,t)$, denoted herein as $\Omega(t)$, are presented in \cite{hook_performance_2018} using a directed acyclic graph $\mathcal{G}(\mathcal{V}, \mathcal{E})$ with graph nodes $\mathcal{V}$ and edges $\mathcal{E}$.
This is \emph{not} the graph of computational node-to-node connections but is instead a directed acyclic graph representation of the evolution of the collective computation: each solution approximation at each computational node (i.e., each $\vec{x}_j^\kappa$) is an element of $\mathcal{V}$, and there is an edge in $\mathcal{E}$ from $\vec{x}_j^\lambda$ to $\vec{x}_i^\kappa$ iff $\edgelong{i}{j}{\kappa} = \lambda$.
Figure \ref{fig:dag} presents a simple example of a directed acyclic graph for the algorithm evolution between two nodes.
The initial states on node $1$ and node $2$ are denoted by $\vec{x}_1^0$ and $\vec{x}_2^0$, respectively.
The initial solution approximation on node $2$ is received by node $1$ and used to compute the next solution approximation on node $1$:
\begin{gather*}
	\vec{x}_1^1 = M_{11}\vec{x}_1^0 + M_{12}\vec{x}_2^0.
\end{gather*}
Node 2, on the other hand, uses $\vec{x}_1^1$ instead of $\vec{x}_1^0$ to compute the next approximation
\begin{gather*}
	\bx_2^1 = M_{22}\bx_2^0 +M_{12}\bx_1^1 ,
\end{gather*}
as depicted in Figure \ref{fig:dag}.
Hook and Dingle \cite[Theorem 1]{hook_performance_2018} prove that the error operator $\Omega(\psi,M,t)$ consists of sums over all the paths within $\mathcal{G}(\mathcal{V}, \mathcal{E})$ of the corresponding iteration matrix blocks.
For example, the error operator at time $t_1$ in Figure \ref{fig:dag} is
\begin{gather*}
	\Omega(\psi,M,t_1 )= \begin{bmatrix}
		M_{11}M_{11}M_{11}+ M_{12}M_{21}M_{11} & M_{12}M_{22}+M_{11}M_{12} \\
		M_{22}M_{21}M_{11}+M_{21}M_{11} & M_{22}M_{22}+M_{21}M_{12} +M_{22}M_{21}M_{12}
	\end{bmatrix}.
\end{gather*}
The authors further show that the convergence rate is bounded by the slowest propagation of information, defined as the shortest path in $\mathcal{G}(\mathcal{V}, \mathcal{E})$ from an initial solution approximation to a current approximation, leading to the error bound that forms the basis of our rejection criteria.
Given a non-negative iteration matrix $M$, i.e., all elements of $M$ are non-negative, Hook and Dingle \cite[Theorem 3]{hook_performance_2018} show that $\Omega(t)$ is bounded as follows
\begin{gather}
  \|\Omega(t)\|_2 \leq \left\|\sum_{k=s(t)}^{l(t)} M^k \right\|_2,
  \label{eq:error-bound}
\end{gather}
where $s(t)$ and $l(t)$ are the lengths of the shortest and longest paths in $\mathcal{G}(\mathcal{V},\mathcal{E})$ at time $t$, respectively.
The goal now is to use \eqref{eq:error-bound} to develop a criterion for whether computational node $i$ should accept or reject a new solution approximation $\vec{x}_j^{\edgelong{i}{j}{\nu_i(t)}}$ obtained from node $j$.
For notational brevity, we introduce $\edge{i}{j}{\kappa} := \edgelong{i}{j}{\kappa}$.

To compare the new solution approximation $\vec{x}_j^{\edge{i}{j}{\nu_i(t)}}$ to the previous solution approximation received by computational node $i$ from computational node $j$, we define $\zeta_{ij}(t)$ to be the index of the solution approximation on node $i$ that was last directly influenced by a solution approximation from node $j$.
In other words, we seek to denote the two most recent solution approximations received by node $i$ from node $j$ at time $t$ as $\vec{x}_j^{\edge{i}{j}{\nu_i(t)}}$ and $\vec{x}_j^{\edge{i}{j}{\zeta_{ij}(t)}}$, respectively.
Formally, $\zeta_{ij}(t) = \arg \max_\kappa \{ \edge{i}{j}{\kappa} < \edge{i}{j}{\nu_i(t)}\}$.
Now, a bound on $\|\vec{x}_{j}^{\edge{i}{j}{\nu_i(t)}} - \vec{x}_{j}^{\edge{i}{j}{\zeta_{ij}(t)}}\|_2$ can be derived using \eqref{eq:error-bound}.
Let $\tau_{ij} [\kappa]$ be the time at which the solution approximation existed on computational node $j$ that would later be used to form $\vec{x}_i^\kappa$, so that $\nu_j \big(\tau_{ij}[\zeta_{ij}(t)]\big) = \edge{i}{j}{\zeta_{ij}(t)}$ and $\nu_j \big(\tau_{ij}[\nu_i(t)]\big) = \edge{i}{j}{\nu_i(t)}$.
Note that $\vec{x}_j^{\edge{i}{j}{\nu_i(t)}}$ can now be expressed as $\vec{x}_j^{\nu_j(\tau_{ij}[\nu_i(t)])} = \vec{\widetilde{x}}_j\big(\tau_{ij}[\nu_i(t)]\big)$ and $\vec{x}_j^{\edge{i}{j}{\zeta_{ij}(t)}}$ as $\vec{x}_j^{\nu_j(\tau_{ij}[\zeta_{ij}(t)])} = \vec{\widetilde{x}}_j\big(\tau_{ij}[\zeta_{ij}(t)]\big)$.
Thus, the following bound holds
\begin{gather*}
  \vec{x}_j^{\edge{i}{j}{\nu_i(t)}}  - \vec{x}_j^{\edge{i}{j}{\zeta_{ij}(t)}}
    = \underbrace{\vec{\widetilde{x}}_j\big(\tau_{ij}[\nu_i(t)]\big) - \vec{x}^*_j}_{\left[\Omega\big(\tau_{ij}[\nu_i(t)]\big)\vec{e}(0)\right]_j}
      + \underbrace{\vec{x}^*_j - \vec{\widetilde{x}}_j\big(\tau_{ij}[\zeta_{ij}(t)]\big)}_{\left[-\Omega\big(\tau_{ij}[\zeta_{ij}(t)]\big)\vec{e}(0) \right]_j} \\
  \Rightarrow \|\vec{x}_j^{\edge{i}{j}{\nu_i(t)}}  - \vec{x}_j^{\edge{i}{j}{\zeta_{ij}(t)}}\|_2
    \leq \bigg [ \big\|\Omega\big(\tau_{ij}[\zeta_{ij}(t)]\big)\big\|_2 +  \big\|\Omega\big(\tau_{ij}[\nu_i(t)]\big) \big\|_2 \bigg] \|\vec{e}(0)\|_2
\end{gather*}
Assuming the initial solution approximation is the zero vector, one has $\|\vec{e}(0)\|_2 \leq \|A^{-1}\|_2 \|\vec{b}\|_2$.
Assuming also that the iteration matrix $M$ is non-negative, the Hook and Dingle bound \eqref{eq:error-bound} is now applied to obtain
\begin{gather}
  \|\vec{x}_j^{\edge{i}{j}{\nu_i(t)}}  - \vec{x}_j^{\edge{i}{j}{\zeta_{ij}(t)}}\|_2 \leq \bigg[ \left \| \sum_{k=s(\tau_{ij}[\zeta_{ij}(t)])}^{l(\tau_{ij}[\zeta_{ij}(t)])} M^k\right\|_2 + \left\| \sum_{k=s(\tau_{ij}[\nu_i(t)])}^{l(\tau_{ij}[\nu_i(t)])} M^k \right \|_2 \bigg ] \|A^{-1}\|_2 \|\vec{b}\|_2.
  \label{eq:rejection-theoretical}
\end{gather}

Evaluating the bound \eqref{eq:rejection-theoretical} directly is very difficult in practice, primarily because none of $A^{-1}$, the $\tau_{ij}$ map, nor the $s(t)$ and $l(t)$ functions are known \textit{a priori}.
Thus, to obtain a practical version of \eqref{eq:rejection-theoretical}, the two individual finite series are bounded by a single infinite series
\begin{gather*}
  \|\vec{x}_j^{\edge{i}{j}{\nu_i(t)}}  - \vec{x}_j^{\edge{i}{j}{\zeta_{ij}(t)}}\|_2 \leq 2 \|A^{-1}\|_2 \|\vec{b}\|_2 \sum_{k=s(\tau_{ij}[\zeta_{ij}(t)])}^\infty \|M\|_2^k.
\end{gather*}
Recall that $\| M \|_2$ is equal to the largest singular value of $M$, denoted as $\sigma_{\max}(M)$, and that $\| A^{-1} \|_2$ is equal to the reciprocal of the smallest singular value of $A$, denoted as $1/\sigma_{\min}(A)$.
Finally, introduce $\tilde{s}_i(t)$ as a lower bound on $\min_{r\neq i} s\big(\tau_{ir}[\zeta_{ir}(t)]\big)$, so that if the geometric series above converges (i.e., if $\|M\|_2 < 1$), then
\begin{gather}
  \|\vec{x}_j^{\edge{i}{j}{\nu_i(t)}}  - \vec{x}_j^{\edge{i}{j}{\zeta_{ij}(t)}}\|_2 \leq 2 \, \frac{\|\vec{b}\|_2}{\sigma_{\min}(A)} \, \frac{\sigma_{\max}(M)^{\tilde{s}_i(t)}}{1 - \sigma_{\max}(M)}.
  \label{eq:rejection}
\end{gather}

The lower bound $\tilde{s}_i(t)$ is obtained in the following manner: each computational node $r$ sends its current value of $\tilde{s}_r(t)$ along with the current solution approximation to its neighbors.
When computational node $i$ receives a value of $\tilde{s}_r(t)$ from node $r$, that value is stored by node $i$ as $\tilde{s}_r$.
Additionally, every time node $i$ computes a new solution approximation, a separate counter $\tilde{s}_i^0$ is incremented.
Once computational node $i$ has received a value from each neighbor $r$ with $M_{ir} \neq 0$, the values for both $\tilde{s}_i(t)$ and $\tilde{s}_i^0$ are set to $\min\left( \tilde{s}_i^0,\; 1+\min_{r: M_{ir} \neq 0} \tilde{s}_r \right)$.
Then the process repeats, with computational node $i$ again collecting updated values for all relevant $\tilde{s}_r(t)$ before updating $\tilde{s}_i(t)$.
Note that since the value received from computational node $j$ for $\tilde{s}_j(t) + 1$ should never be less than $\tilde{s}_i(t)$, the solution approximation $\vec{x}_j^{\edge{i}{j}{\nu_i(t)}}$ will only be accepted by computational node $i$ if \eqref{eq:rejection} is satisfied and the new value of $\tilde{s}_j(t)$ is such that $\tilde{s}_j(t) + 1 \geq \tilde{s}_i(t)$.
This additional constraint provides some resilience for when the value of $\tilde{s}_j(t)$ is itself corrupted.
These two constraints form the rejection criterion for the rejection variant of the asynchronous Jacobi method presented in Algorithm~\ref{alg:rejection}.
\begin{algorithm2e}[H] \label{alg:rejection}
  \ForEach{node i=1,2,\ldots,N}
  {
    Initialize the algorithm with $\vec{x}_i^0 = \vec{0}$, $\tilde{s}_i = 0$, and $\tilde{s}_i^0 =0$. Set $\kappa=0$ and $\mathcal{S}=\{\}$. \\
    \ForEach{$\vec{x}_j$ and $\tilde{s}_j$ received from node $j$}
    {
      \If{$\|\vec{x}_j - \vec{x}_j^\kappa\|_2 \leq 2 \frac{\|b\|_2}{\sigma_{\min}(A)} \frac{\sigma_{\max}(M)^{\tilde{s}_i}}{1 - \sigma_{\max}(M)}$ and $\tilde{s}_j+1 \geq \tilde{s}_i$}
      {
        set $\vec{x}_j^\kappa = \vec{x}_j$ \\
        store $\tilde{s}_j$ in $\mathcal{S}$ \\
        \If{$\mathcal{S}$ contains $\tilde{s}_r$ for all $r$ such that $M_{ir} \neq 0$}
        {
          set $\tilde{s}_i = \min \{ \tilde{s}_i^0, 1 + \min \mathcal{S} \}$ \\
          set $\tilde{s}_i^0 = \tilde{s}_i$ \\
          set $\mathcal{S} = \{\}$
        }
        set $\vec{x}_i^{\kappa+1} = \sum_{r=1}^N M_{ir} \vec{x}_r^\kappa + \vec{c}_i$ \\
        set $\tilde{s}_i^0 = \tilde{s}_i^0 + 1$ \\
        communicate $\vec{x}_i^{\kappa+1}$ and $\tilde{s}_i$ \\
        set $\kappa = \kappa + 1$
      }
    }
  }
  \caption{Asynchronous Jacobi Rejection Variant (ASJ-R)}
\end{algorithm2e}

The development of stopping criteria for asynchronous methods remains an active area of research.
Hook and Dingle \cite{hook_performance_2018} have each node report to a root node when a local stopping criterion is met.
Each node then continues iterating until the root node indicates that it has received reports from all nodes.
Instead of designating a root node, our approach has each node collect such reports in a decentralized fashion.
Each node $i$ checks the following criterion after a new solution approximation $\vec{x}_i^{\nu_i(t)}$ is computed:
\begin{gather}
  \| D_{ii} d\vec{x}_i^{\nu_i(t)} \|_\infty < \epsilon \frac{\|b\|_2}{\sqrt{m}},
  \label{eq:stopping}
\end{gather}
where $d\vec{x}_i^{\nu_i(t)} = \vec{x}_i^{\nu_i(t)} - \vec{x}_i^{\nu_i(t)-1}$ and $\epsilon>0$ is a prescribed tolerance.
If the bound in \eqref{eq:stopping} is satisfied, node $i$ reports to all other nodes that it has locally converged while continuing to iterate.
If a successive solution approximation on node $i$ fails to satisfy \eqref{eq:stopping}, node $i$ reports to all other nodes that it has no longer locally converged.
If a node $j$ has locally converged and has received reports that all other nodes have locally converged, it starts a ``convergence duration'' timer while continuing to iterate.
If node $j$ either determines it has no longer locally converged or receives a report that another node has no longer locally converged, the timer is set back to zero.
Each node continues to iterate until either (i) the specifying convergence duration is achieved or (ii) a specified maximum number of iterations is reached.
Ideally, the convergence duration is chosen so that the information that a given node has no longer locally converged has time to arrive at all other nodes before those other nodes stop iterating.
As such, the proper choice of convergence duration likely depends on parameters that determine the speed at which information propagates across the nodes, e.g., network hardware, sparsity pattern of $A$, etc.

\section{Numerical Results}
\label{sec:results}
Having derived the modified asynchronous Jacobi in Section \ref{sec:modification}, shown in Algorithm \ref{alg:rejection}, we now numerically evaluate the proposed method.
We choose the benchmark problem to have an analytic solution so we can verify the implementation of Algorithm~\ref{alg:rejection}.
We then compare the method convergence against that of the traditional asynchronous Jacobi method when the natural and malevolent data corruption described in Sections \ref{sec:natural-corruption} and \ref{sec:malevolent-corruption}, respectively, are present.
The comparison includes results for both when the singular values in the path-length rejection criterion \eqref{eq:rejection} are ``exact'' (computed by each node before the iteration using an eigensolver) and when the singular values are approximated with varying levels of relative accuracy.

Each run is to a convergence tolerance of $\epsilon = 10^{-5}$ and performed on a single 36-core node of the Quartz supercomputer at the Livermore Computing Complex.
The ``exact'' singular values in the rejection criterion \eqref{eq:rejection} are computed using the bi-diagonal divide and conquer singular value decomposition algorithm in the \emph{Eigen} software library \cite{eigen}.
The \emph{Skywing} software platform \cite{Skywing}, which is designed to support asynchronous algorithms, is used for execution and communication for all methods.
Each run leverages $16$ \emph{Skywing} agents to serve as the $N=16$ computational nodes.
More on the motivation and implementation of \emph{Skywing} can be found in \cite{erlandson_resilient_2023} and on GitHub at \url{https://github.com/LLNL/Skywing}, respectively.

To account for the stochastic nature of both asynchronous algorithms and the corruption model, ensembles of $30$ runs are performed for each study.
Quantities such as the ensemble wall-clock time and relative solution error are reported as a geometric mean defined as
\begin{gather*}
\bar{a} = \exp \left( \frac{1}{s} \sum_{k=1}^s \log \big(a_k\big) \right),
\end{gather*}
where $a$ is the quantity of interest and $s=30$ corresponds to the ensemble of $30$ runs.
When $a$ is time-dependent, such as when $a$ represents the relative solution error $\|\vec{e}(t)\|_2 / \|\vec{x}^*\|_2$, the values of $\bar{a}(t)$ are obtained using linear interpolants of $a_k(t)$.
If $a_k(t)$ contains an \textsc{ieee} 754 \texttt{NaN} or $\pm \infty$ value, as can happen in the relative solution error with data corruption, the corresponding value of $\bar{a}(t)$ is omitted.

The linear system \eqref{eq:axb} solved throughout this section is obtained from a finite difference discretization of the following Poisson problem on the unit square
\begin{gather}
  \begin{gathered} \label{eq:poisson}
	   -\left( \frac{\partial^2 u}{\partial x^2} + \frac{\partial^2 u}{\partial y^2} \right) = f,
      \quad x \in (0,1), y \in (0,1), \\
      \quad u(0,y)=u(1,y)=u(x,0)=u(x,1)=0,
  \end{gathered}
\end{gather}
where the choice of $f(x,y)=2\pi^2 \sin(\pi x) \sin(\pi y)$ results in an analytic solution $u(x,y) = \sin(\pi x) \sin(\pi y)$.
The unit square is uniformly discretized into $\ell +1 \times \ell+1$ squares of length $h=1/(\ell+1)$.
Such a discretization along with the Dirichlet boundary condition in \eqref{eq:poisson} leaves the values of $u(x_i, y_j)$ to be determined at the square vertices, where $x_i = (i+1)h$ and $y_j=(j+1)h$ for $i=0,\ldots,\ell-1$ and $j=0,\ldots,\ell-1$.
Let the $k$-th element of $\vec{x} \in \mathbb{R}^{\ell^2}$ and $\vec{b} \in \mathbb{R}^{\ell^2}$ in \eqref{eq:axb} be $u(x_i, y_j)$ and $f(x_i, y_j)$, respectively, with $i = (k \mod \ell)$ and $j = k \ell$.
With the Laplace operator discretized across the points $(x_i,y_j)$ using centered finite difference, the matrix $A$ in \eqref{eq:axb} is defined as the following $\ell^2 \times \ell^2$ block tridiagonal matrix
\begin{gather*}
  A=
    \begin{bmatrix}
       L & -I     &        &        &    \\
      -I &  L     &     -I &        &    \\
         & \ddots & \ddots & \ddots &    \\
         &        &     -I &      L & -I \\
         &        &        &     -I &  L
    \end{bmatrix}
  \text{, where }
  L =
    \begin{bmatrix}
           4 & -1     &        &        &    \\
          -1 &  4     &     -1 &        &    \\
             & \ddots & \ddots & \ddots &    \\
             &        &     -1 &      4 & -1 \\
             &        &        &     -1 & 4
    \end{bmatrix},
\end{gather*}
and $I \in \R^{\ell \times \ell}$ is the identity matrix.
The linear system is evenly distributed across the agents, i.e., $m_1=\ldots=m_16$.

\subsection{Convergence Duration}
\label{sec:convergence_duration}

\begin{figure}[htbp]
    \centering
    \includegraphics[width=0.32\textwidth]{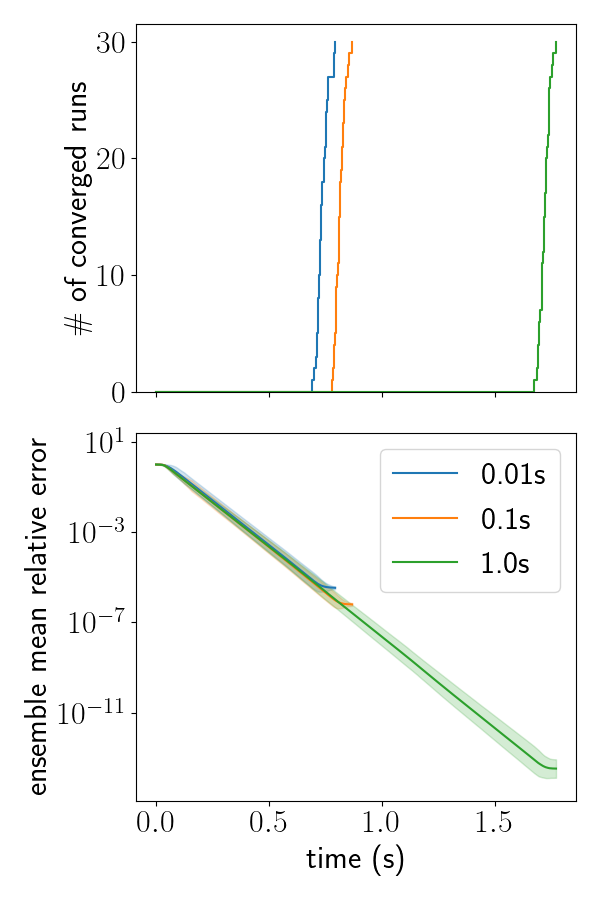}
    \includegraphics[width=0.32\textwidth]{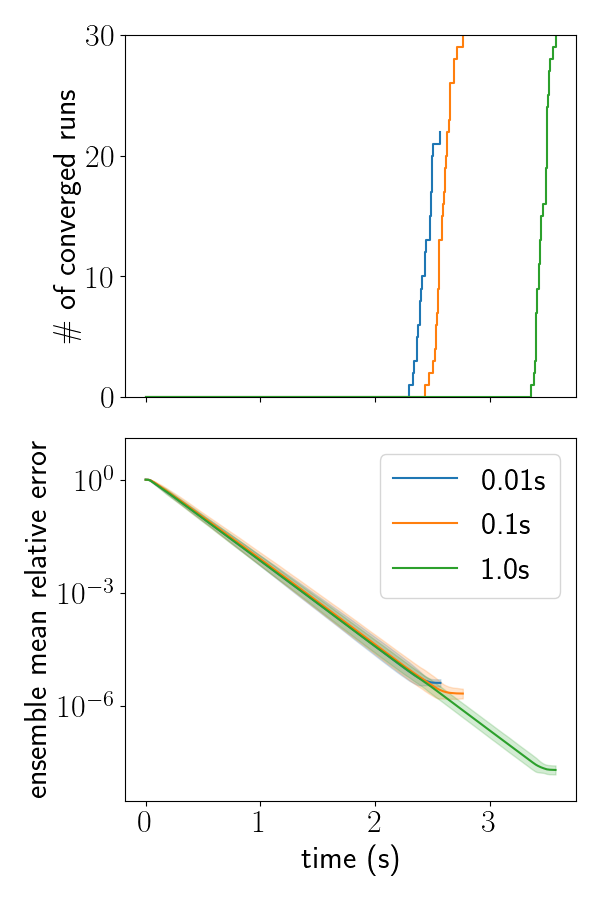}
    \includegraphics[width=0.32\textwidth]{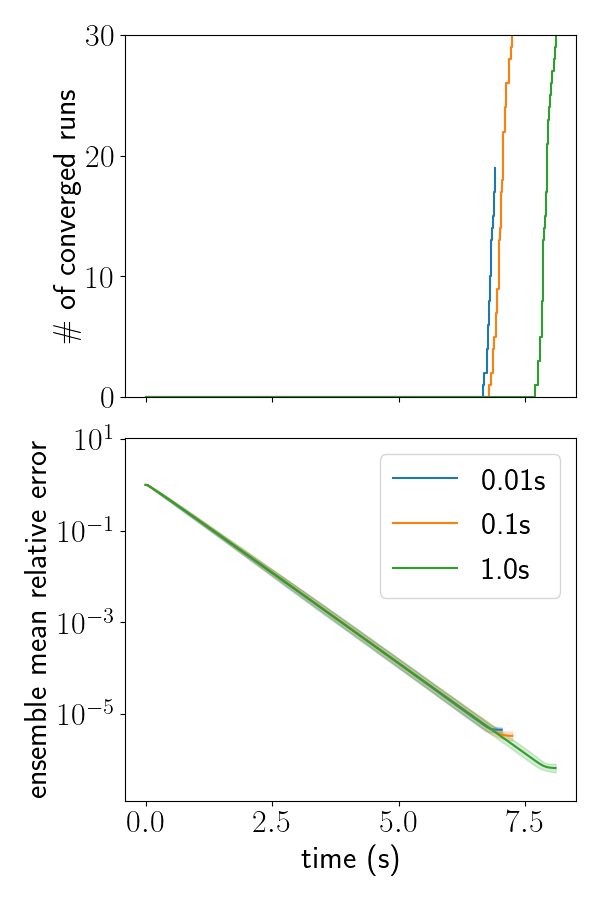}
    \caption{
      Ensemble convergence of ASJ-R for various convergence durations on Poisson benchmark problem with $m=144$ (left), $m=400$ (center), and $m=784$ (right).
      All ensemble runs converge for all durations with the smallest system size; however, the larger system sizes indicate the agents are unable to reach consensus on convergence for the smallest duration.
    }
    \label{fig:duration}
  \end{figure}

  Recall that the stopping criteria discussed at the end of Section \ref{sec:modification} involves a user-specified convergence duration for both the traditional asynchronous Jacobi (ASJ) and the modified asynchronous Jacobi (ASJ-R) methods.
  As such, we first evaluate the impact of the convergence duration on the asynchronous solving of the benchmark problem \eqref{eq:poisson} in the absence of data corruption.
  The number of squares chosen to discretize the unit square domain is selected so the linear system results in $m=144$, $m=400$, and $m=784$ unknowns, respectively.
  Figure \ref{fig:duration} shows the dependence of the convergence behavior on the convergence duration.
  For the smallest system, $m=144$, all runs converge for all three duration values both in the sense that the agents reach consensus on convergence and that the resulting relative solution error is below the target tolerance.
  As the system size increases to $m=400$, however, the shortest duration $0.01$s results in around $30\%$ of the ensemble runs ending with the agents unable to reach consensus on convergence, even though the relative solution error is near or below the target tolerance.
  With the largest system size $m=784$, the shortest duration results in slightly more than $30\%$ of the ensemble runs failing to reach consensus.
  Increasing the duration to $0.1$s remedies the lack of consensus for all three system sizes evaluated.
  While these results could motivate a duration of $0.1$s for evaluating the methods with corruption present, we choose the $1$s duration that is an order of magnitude longer to increase confidence that convergent results are due to the algorithm being able to continue decreasing the error despite corruption and not the iteration stopping at the right moment between corruption occurrences.

\subsection{Path Length Rejection Variant with Data Corruption}
\label{sec:rejection-results}

  With the convergence duration established, we now evaluate the resilience of the ASJ and ASJ-R methods to both natural and malevolent corruption, as defined in Section \ref{sec:natural-corruption} and Section \ref{sec:malevolent-corruption}, respectively.
  To choose a system size, we consider the sizes of the power flow benchmark systems mentioned in \cite{wang_distribution_2019}, which range from $\mathcal{O}(14)$ to $\mathcal{O}(300)$ unknowns.
  As such, all corruption studies discretize the unit square such that the linear system has $m=400$ unknowns ($\ell=20$) and results in each agent communicating with one or two neighbors in a line configuration.
  We present time for the corruption studies as relative to the average time to convergence ($\approx 3.5$s) obtained in Section \ref{sec:convergence_duration} for $m=400$, i.e., a time to convergence of $1.0$ indicates the method converged in the same amount of time as it would without corruption present.

  \subsubsection*{Natural Data Corruption}

    Our first investigation introduces bit flips to communicated data, similar to the studies performed by Anzt et al. in \cite{anzt_2015}.
    We aim to assess the impact of the probability $p$ of a bit flip on the convergence of both ASJ and ASJ-R.
    As discussed in Section \ref{sec:natural-corruption}, corruption is applied at each iteration and on every agent with probability $p$ to all communicated data.
    The elements of $\vec{x}_i^\kappa$ in both ASJ and ASJ-R are stored as \textsc{ieee} 754 double floating point numbers, whereas the values of the approximate shortest path length $\tilde{s}_i(t)$ in ASJ-R are stored as signed integers.
    If a given double floating point value is chosen to be corrupted, a bit index out of a given subset of its $64$ bit representation is randomly chosen to be flipped.
    If a given signed integer value is chosen to be corrupted, a bit index out of any of its $32$ bit representation is randomly chosen to be flipped.

\begin{figure}[htbp]
      \centering
      \includegraphics[width=0.45\textwidth]{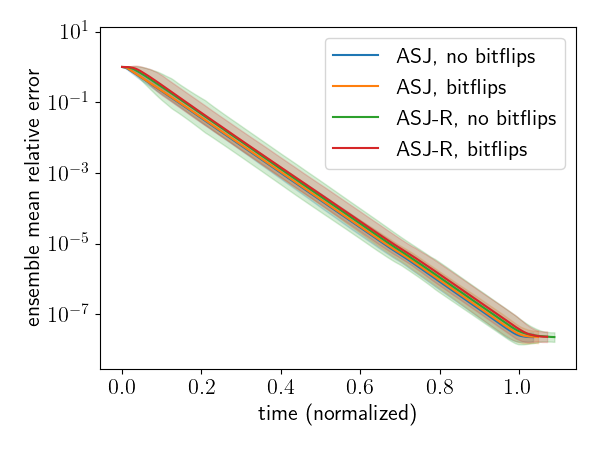}
      \includegraphics[width=0.45\textwidth]{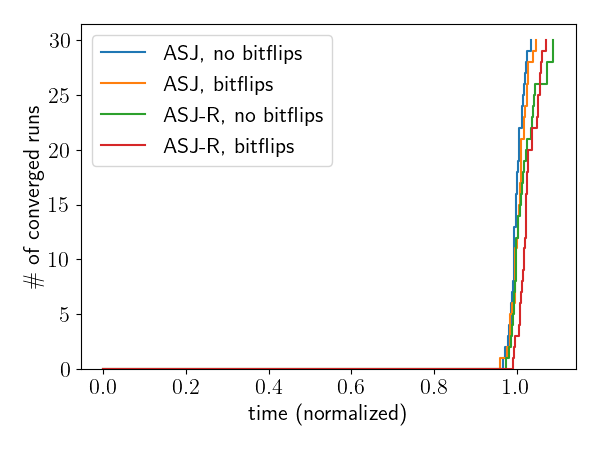}
      \caption{
        Ensemble convergence of ASJ and ASJ-R with bit flip probability $p=0.01$, with double floating point flips limited to the lower mantissa \bitsLowerMantissa{}.
        Convergence is achieved in all ASJ and ASJ-R runs, with times to solution comparable to the respective baseline (no corruption) values.
      }
      \label{fig:bitflip-lower-mantissa}
    \end{figure}

    For our first study, we fix the probability of a bit flip in a given communicated value to be $p=0.01$.
    Following Anzt et al. \cite{anzt_2015}, we investigate the following double floating point subsets: the lower mantissa \bitsLowerMantissa{}, the upper mantissa \bitsUpperMantissa{}, the exponent \bitsExponent{}, and the sign bit \bitsSign{}.
    We start with the lower mantissa subset \bitsLowerMantissa{}, which leads to floating point value corruption ranging from $1/2^{52} \approx 10^{-16}$ to $1/2^{27} \approx 10^{-8}$ relative to the original values.
    Figure~\ref{fig:bitflip-lower-mantissa} shows the convergence behavior for ASJ and ASJ-R.
    For both ASJ and ASJ-R, all runs in the respective bit flip ensemble converge with times to solution that are approximately the same as those of the respective baseline (no corruption) ensemble.
    The indifference of the ASJ convergence behavior to lower-mantissa flips is consistent with Anzt et al. \cite{anzt_2015}, where it was found that ASJ convergence behavior is not affected by such bit flips until the relative residual norm is reduced to a very small value.
    The indifference is due to the corruption caused by lower-mantissa flips being too small to significantly affect the iteration evolution to the tolerance $\epsilon=10^{-5}$, as relaxation methods are inherently robust to small amounts of corruption.

\begin{figure}[htbp]
      \centering
      \includegraphics[width=0.45\textwidth]{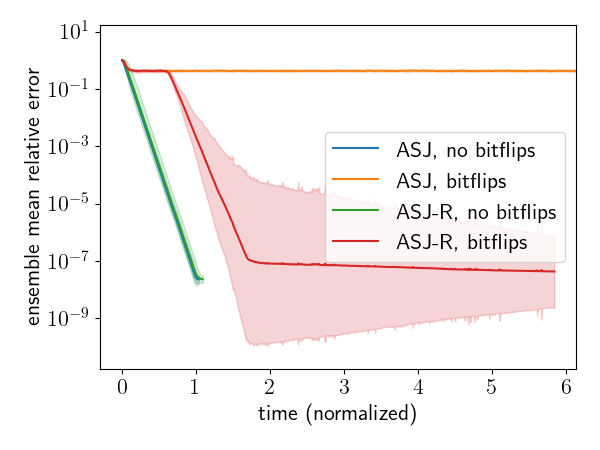}
      \includegraphics[width=0.45\textwidth]{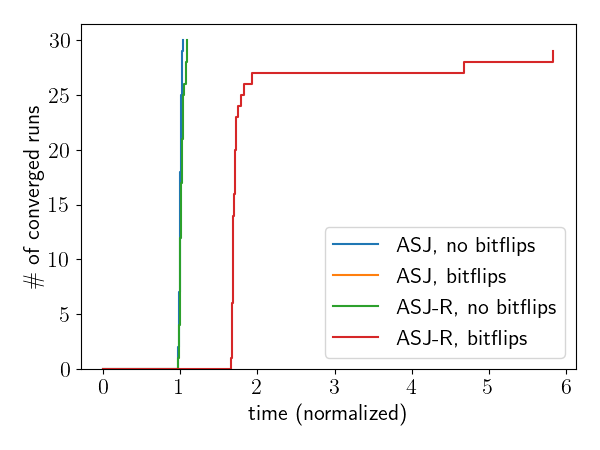}
      \caption{
        Ensemble convergence of ASJ and ASJ-R with bit flip probability $p=0.01$, with double floating point flips limited to the sign bit \bitsSign{}.
        Convergence is lost for all of the ASJ runs and achieved for all ASJ-R runs, albeit with longer times to solution.
      }
      \label{fig:bitflip-sign}
    \end{figure}

\begin{figure}[htbp]
      \centering
      \includegraphics[width=0.45\textwidth]{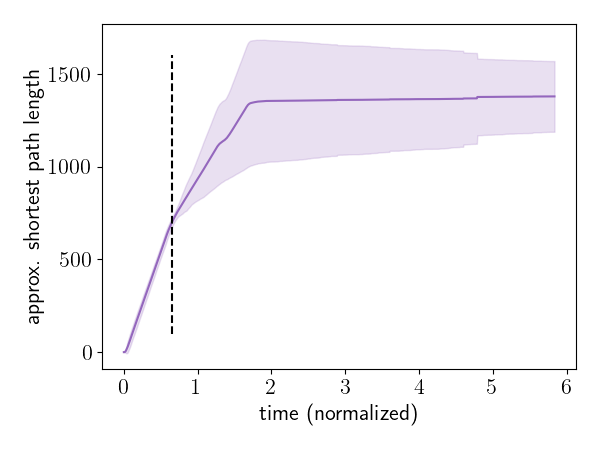}
      \includegraphics[width=0.45\textwidth]{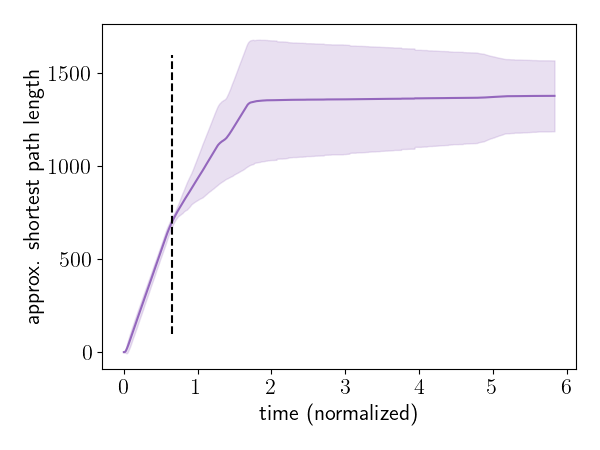}
      \caption{
        Approximate shortest path length $\tilde{s}_i(t)$ used by ASJ-R algorithm with bit flip probability $p=0.01$, with double floating point flips limited to the sign bit \bitsSign{} (left: $i=7$, right: $i=9$).
        The approximate shortest path length reaches $700$ at around the time the stagnation period ends in Figure~\ref{fig:bitflip-sign} (denoted by dashed black line).
      }
      \label{fig:bitflip-sign-length}
    \end{figure}

    To introduce larger corruption, we now investigate the sign bit \bitsSign{} subset, which leads to floating point value corruption of $2$ relative to the original values.
    Figure~\ref{fig:bitflip-sign} shows the convergence behavior for ASJ and ASJ-R.
    For ASJ, the effect of the more significant corruption from sign bit flips is evident as the solution error of all ASJ runs decreases at first but then stagnates at a level well above the convergence tolerance, consistent with the findings of \cite{anzt_2015}.
    The introduction of the rejection criterion in ASJ-R, based on \eqref{eq:rejection}, restores convergence in all of the runs, albeit with longer times to solution: $1.75$x longer for about $80\%$ of the runs and $6$x longer for the slowest run.
    This increased time to solution is explained starting with the presence of a stagnation period for all ASJ-R runs with bit flips.
    Given that the ASJ-R error during this stagnation period coincides with the stagnated ASJ error, it can be inferred that the value of the approximate shortest path length $\tilde{s}_i(t)$ in \eqref{eq:rejection} during the stagnation period is not yet large enough for ASJ-R to reject the corruption.
    Figure~\ref{fig:bitflip-sign-length} shows the value of the approximate shortest path length for two agents.
    The values of $\tilde{s}_7(t)$ and $\tilde{s}_9(t)$ grow roughly linearly with time until the end of the stagnation period around $t=0.6$.
    Around $t=0.65$, the value of $\tilde{s}_i(t)$ is large enough ($\approx 700$) to start rejecting data containing bit flips so that all $30$ runs can resume converging.
    The values of $\tilde{s}_7(t)$ and $\tilde{s}_9(t)$ continue to grow after $t=0.65$, albeit at a slower rate due to the rejections.

    To introduce corruption with a relative magnitude between the sign bit and lower mantissa subsets, we now investigate the upper mantissa \bitsUpperMantissa{} subset, which leads to floating point value corruption ranging from $1/2^{26} \approx 10^{-7}$ to $1/2$ relative to the original values.
    \begin{figure}[htbp]
      \centering
      \includegraphics[width=0.45\textwidth]{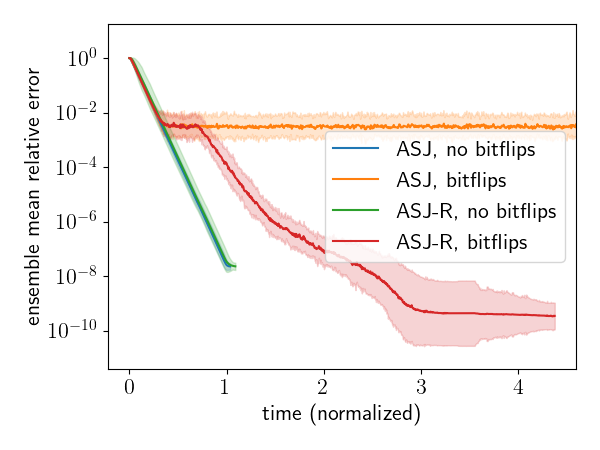}
      \includegraphics[width=0.45\textwidth]{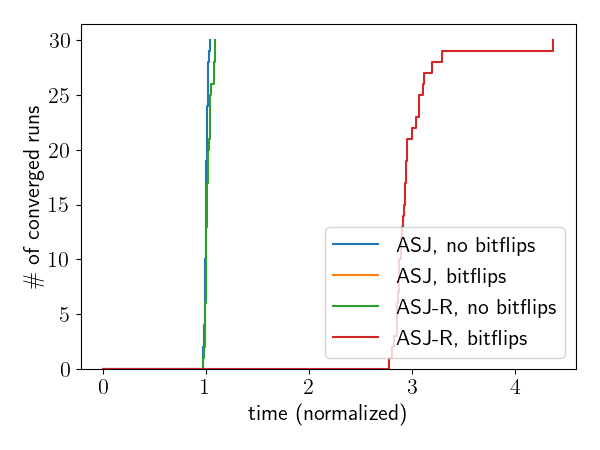}
      \caption{
        Ensemble convergence of ASJ and ASJ-R with bit flip probability $p=0.01$, with double floating point flips limited to the upper mantissa \bitsUpperMantissa{}.
        Convergence is lost for all of the ASJ runs and achieved for all ASJ-R runs. The times to solution of the convergent runs are longer than those of sign bit flips (Figure~\ref{fig:bitflip-sign}); however, the times to tolerance $\epsilon=10^{-5}$ are about the same as sign bit flips.
      }
      \label{fig:bitflip-upper-mantissa}
    \end{figure}
    In Figure~\ref{fig:bitflip-upper-mantissa}, the corruption from upper-mantissa flips is still large enough to prevent convergence in all ASJ runs, with the solution error stagnating at a level above tolerance but lower than with sign bit flips, consistent with the smaller magnitude value changes and with the findings of \cite{anzt_2015}.
    For ASJ-R, convergence is achieved in all runs with a time to solution around $3$x longer for all but one run that took around $4.25$x longer.
    There is a stagnation period followed by a return to convergence behavior, as seen with sign bit flips, only the period lasts longer until about $t=0.7$.
    There is also a significant reduction in convergence rate starting around $t=1.5$ that is not addressed until around $t=2.75$.
    Both of these differences are explained by the observation that as $\tilde{s}_i(t)$ grows, there is always corruption of a certain magnitude that will not be rejected.
    Thus, to reject the larger of the magnitude corruption range that upper-mantissa bit flips can cause, $\tilde{s}_i(t)$ must reach a larger value than with sign bit flips before convergence is restored from the stagnation period.
    This is confirmed in Figure~\ref{fig:bitflip-upper-mantissa-length}, where the value for $\tilde{s}_i(t)$ at $t=0.7$ is about $750$.
    The degraded convergence from $t=1.5$ until $t=2.75$ is explained by the requirement that $\tilde{s}_i(t)$ reach an even larger value before the smaller of the corruption range is rejected.
    It is worth noting that Anzt et al. \cite{anzt_2015} also see a slower rate of convergence for synchronous Jacobi with bit flips for likely the same reason.
    \begin{figure}[htbp]
      \centering
      \includegraphics[width=0.45\textwidth]{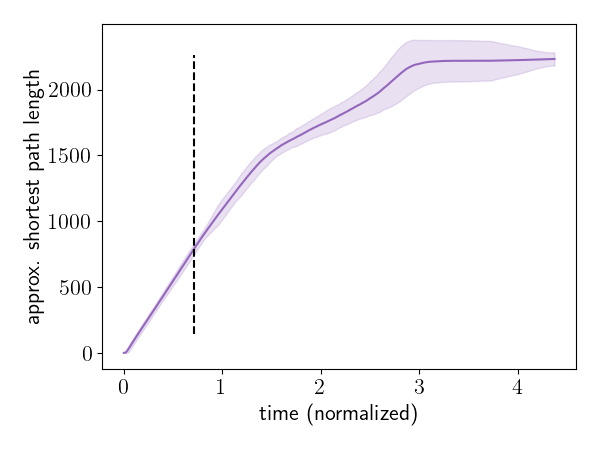}
      \includegraphics[width=0.45\textwidth]{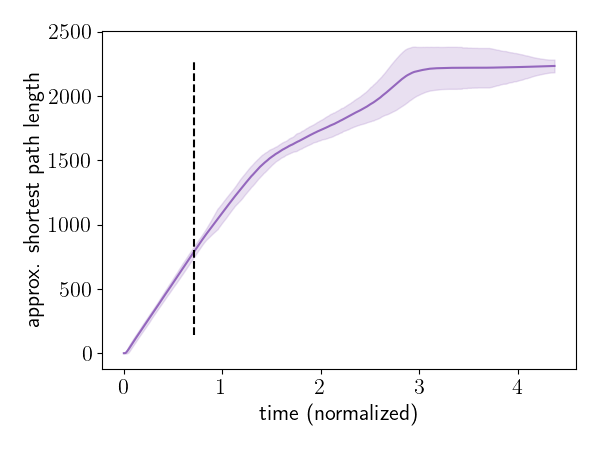}
      \caption{
        Approximate shortest path length $\tilde{s}_i(t)$ used by ASJ-R algorithm with bit flip probability $p=0.01$, with double float point flips limited to the sign bit \bitsUpperMantissa{} (left: $i=7$, right: $i=9$).
        The approximate shortest path length reaches $750$ at around the time the stagnation period ends in Figure~\ref{fig:bitflip-upper-mantissa} (denoted by dashed black line).
      }
      \label{fig:bitflip-upper-mantissa-length}
    \end{figure}

    The last subset to investigate is the exponent subset \bitsExponent{}, which leads to floating point value changes ranging from $1$ to $2^{1023}-1 \approx 10^{308}$ relative to the original values.
    \begin{figure}[htbp]
      \centering
      \includegraphics[width=0.45\textwidth]{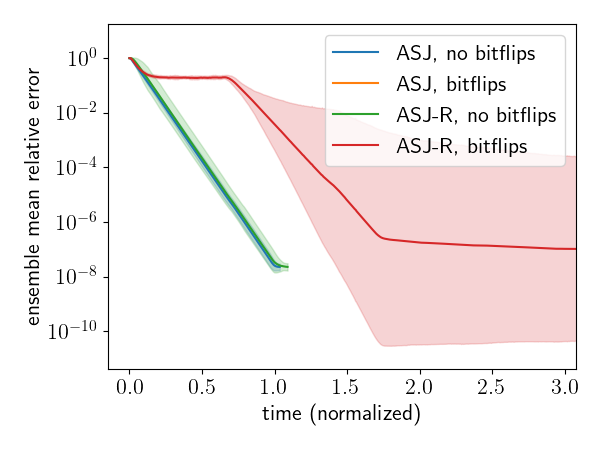}
      \includegraphics[width=0.45\textwidth]{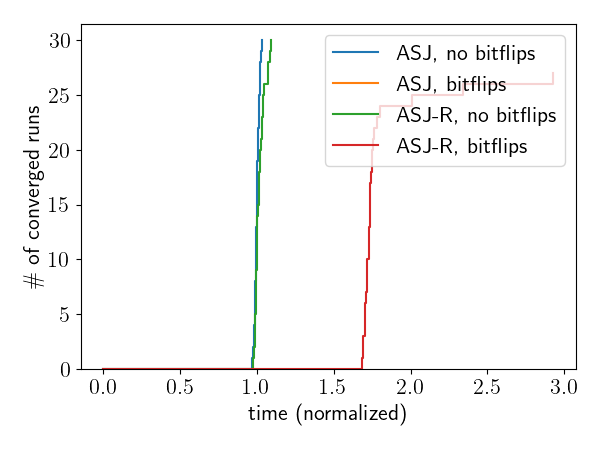}
      \caption{
        Ensemble convergence of ASJ and ASJ-R with bit flip probability $p=0.01$, with double floating point flips limited to the exponent \bitsExponent{}.
        Convergence is lost for all of the ASJ runs and achieved for most of the  ASJ-R runs, with the times to solution being mostly comparable to that of sign bit flips (Figure~\ref{fig:bitflip-sign})
      }
      \label{fig:bitflip-exponent}
    \end{figure}
    In Figure~\ref{fig:bitflip-exponent}, the corruption from exponent flips is large enough that all ASJ runs result in solution iterates containing non-finite (i.e., \textsc{ieee} 754 \texttt{NaN}) values after just a few iterations, which is consistent with the findings of \cite{anzt_2015}.
    For ASJ-R, convergence is achieved in most runs with times to solution similar to that of sign bit flips: around $1.75$x for almost all the runs with one run taking almost $3$x.
    The slowdown to $1.75$x is explained as with the sign and upper-mantissa bit flips: $\tilde{s}_i(t)$ must reach a certain value before larger magnitude corruption data is rejected enough to restore convergence.
    The slowdown to $3$x is explained as with the upper-mantissa bit flips: after the initial stagnation period, $\tilde{s}_i(t)$ must continue to grow before smaller magnitude corruption is also rejected.
    The runs that do not converge are, however, not explained by prior observations for sign or upper-mantissa bit flips.
    In those runs, the corruption that is not rejected during the stagnation period is large enough to cause the evolution of the solution approximations to substantially deviate from that predicted by the convergence theory upon which the rejection criterion \eqref{eq:rejection} is derived.
    The deviation is large enough that the updates required to drive the solution approximation back towards the exact solution are now significantly larger than those predicted by the convergence theory, causing the the rejection criterion to reject the valid updates that would otherwise restore convergence.

\begin{figure}[htbp]
      \centering
      \includegraphics[width=0.45\textwidth]{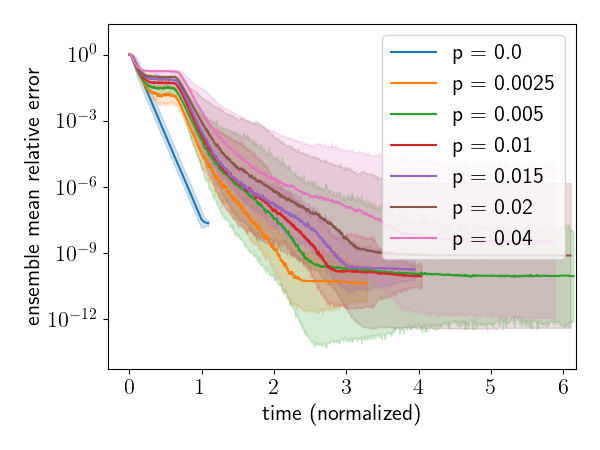}
      \includegraphics[width=0.45\textwidth]{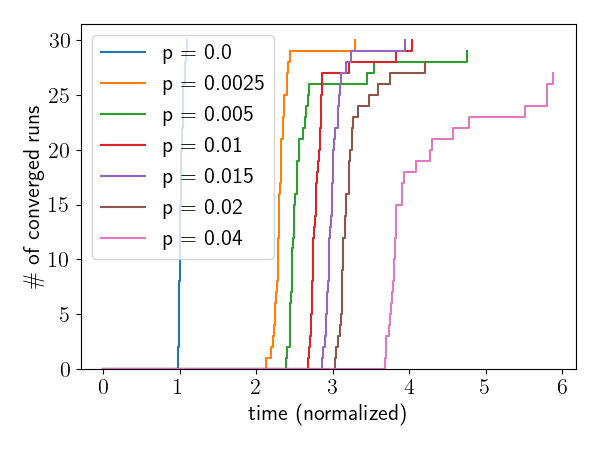}
      \caption{
        Ensemble convergence of ASJ-R with bit flip corruption probabilities ranging from $p=0$ to $p=0.04$, with double floating point flips in any of the $64$ bits.
        Convergence is lost for all of the ASJ runs and achieved for large majority of ASJ-R runs, with increasing $p$ resulting in larger times to solution and larger proportion of ensemble runs that fail to converge.
      }
      \label{fig:bitflip-probabilities}
    \end{figure}

    With an understanding of how ASJ and ASJ-R perform on bit flips with probability $p=0.01$ in subsets of the floating point double, we now investigate flipping any of $64$ bits with probability $p$ values of $0.0025$, $0.005$, $0.01$, $0.015$, $0.02$, and $0.04$.
    All of the runs in any ASJ ensemble corresponding to $p>0$ quickly saw \texttt{NaN} values in the solution approximations.
    Noting similar occurrence of \texttt{NaN} values in Figure~\ref{fig:bitflip-exponent}, one can infer that, even with bit flip probability as low as $p=0.0025$, the ASJ runs quickly experience the occurrence of one or more exponent bit flips.
    For ASJ-R, the convergence behavior is shown in Figure~\ref{fig:bitflip-probabilities}.
    All but one of the runs for $p \leq 0.015$ converged, with a large majority of runs still converging for $p = 0.02$ and $p = 0.04$.
    As one might expect, increasing $p$ results in the times to solution increasing from around $2.25$x to around $4$x, with increasing variability as $p$ increases.
    Such behavior is explained by the prior observations: that increasing $p$ results in increased likelihood of bit flips causing corruption that is small enough to avoid rejection by the increasing $\tilde{s}_i(t)$, yet large enough to (i) delay convergence until $\tilde{s}_i(t)$ increases enough or (ii) cause the solution approximation to deviate substantially enough that the rejection criterion actually prevents convergence.
    All in all, the ASJ-R algorithm has a very high probability of converging even when a large number of bit flips occur, e.g., $p=0.04$ of communicated data are corrupted at each iteration.

  \subsubsection*{Malevolent Data Corruption}

    Our second investigation introduces malevolent manipulation of stored data, as defined in Section \ref{sec:malevolent-corruption}.
    As opposed to the investigation with natural bit flips, here we limit the corruption to double floating point values (i.e., no corruption of signed integer values for ASJ-R).
    We aim to assess the impact of the recovery time $\omega_r$ and mean manipulation offset $\delta$ on the convergence of both ASJ and ASJ-R.
    As described in Section \ref{sec:malevolent-corruption}, while agent $i$ is in a degraded state, every element of $\vec{x}_i^\kappa$ is manipulated by an additive offset sampled from a normal distribution with mean $\delta$ and standard deviation $\tfrac{1}{2}\delta$.
    We introduce corruption to agent $i=9$ with a time-to-failure $\omega_f = 2$s so the first degraded state occurs before the agents would otherwise start the convergence duration timers without corruption ($\approx 2.5$s).
    We study recovery times $\omega_r$ selected from $0.01$s, $0.02$s, $0.03$s, $0.04$s, and $0.05$s that represent relative uptimes of $99.5$\%, $99$\%, $98.5$\%, $98$\%, and $97.5$\%.
    The offset magnitudes $\delta$ are selected from $0.1$, $0.2$, $0.3$, $0.4$, and $0.5$.

\begin{figure}[htbp]
      \centering
      \includegraphics[width=0.45\textwidth]{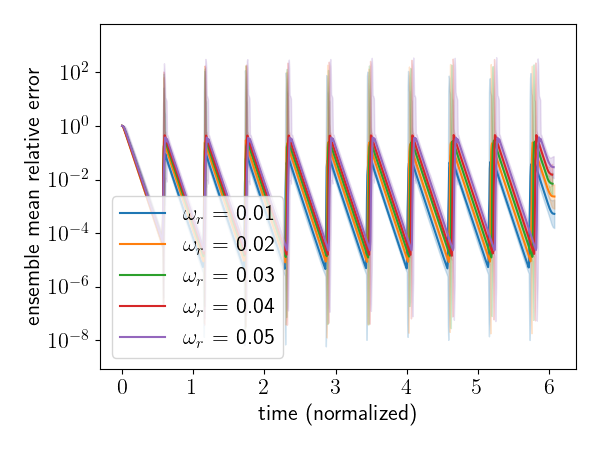}
      \includegraphics[width=0.45\textwidth]{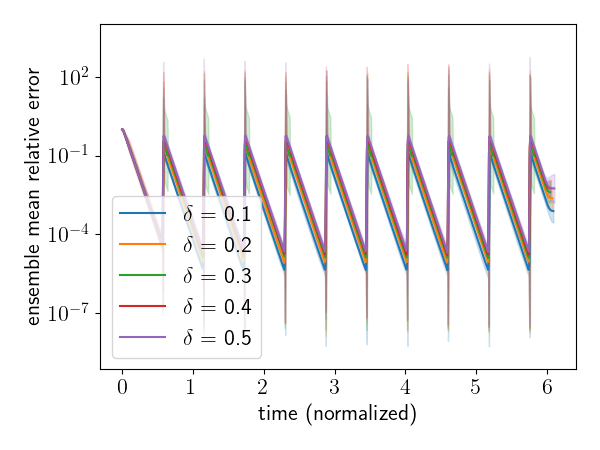}
      \caption{
        Ensemble convergence of ASJ with malevolent corruption with time-to-failure $\omega_f=2$s ($t \approx 0.6$ in normalized time), various recovery times $\omega_r$ and various offset magnitudes $\delta$ (left: various $\omega_r$ values and $\delta=0.2$, right: $\omega_r=0.02$s and various $\delta$ values).
        Convergence is not obtained for any of the runs.
      }
      \label{fig:gaussian-original}
    \end{figure}

    Figure~\ref{fig:gaussian-original} shows the convergence behavior of ASJ for time-to-failure $\omega_f = 2$s with various $\omega_r$ and fixed $\delta=0.2$ and with fixed $\omega_r=0.02$s ($99\%$ uptime) and various $\delta$.
    All of the ASJ runs fail to converge for the recovery times and offset magnitudes explored.
    The effect of the corruption is seen around $t=0.6$ as the solution error is small enough to begin the convergence duration but then rapidly increases due to the corruption on agent $9$ that quickly propagates to other agents.
    The error increases to a peak that coincides with agent $9$ returning to a normal state, after which the error does decrease until the next rapid increase when agent $9$ is again degraded.
    Increasing either the recovery time $\omega_r$ or the offset magnitude $\delta$ effectively shifts the overall error evolution upward.
    While one might technically obtain convergence to the tolerance $\epsilon = 10^{-5}$ by decreasing the convergence duration for the smallest $\omega_r$ and $\delta$, the results in Figure~\ref{fig:gaussian-original} show that the ASJ method cannot reliably converge to a given tolerance with malevolent corruption.

\begin{figure}[htbp]
      \centering
      \includegraphics[width=0.45\textwidth]{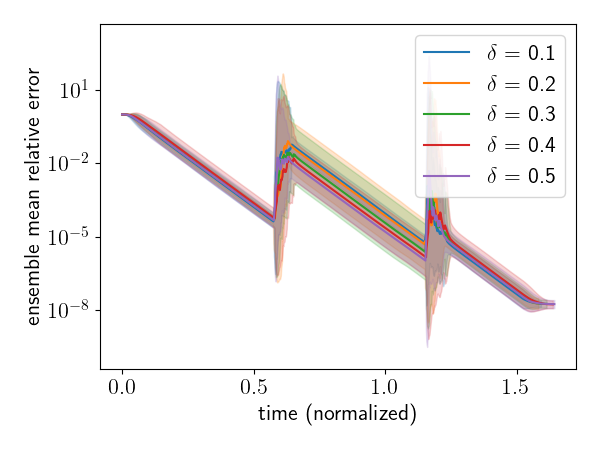}
      \includegraphics[width=0.45\textwidth]{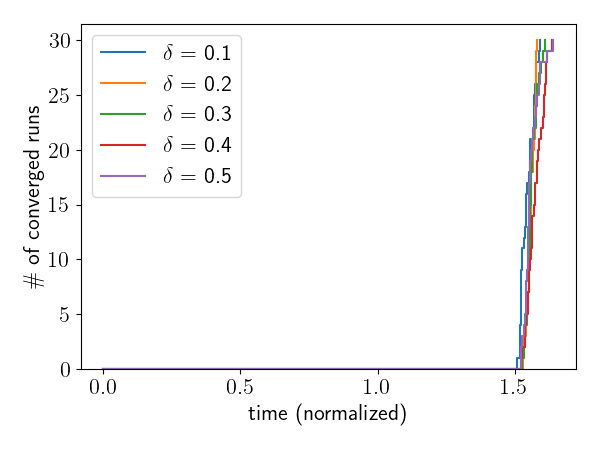}
      \caption{
        Ensemble convergence of ASJ-R with malevolent corruption with time-to-failure $\omega_f=2$s ($t \approx 0.6$ in normalized time), recovery time $\omega_r=0.02$s, and various offset magnitudes $\delta$.
        Convergence is achieved for all runs, with the time to solution being comparable to that of bit flip (Figure~\ref{fig:bitflip-sign}) and exponent flip (Figure~\ref{fig:bitflip-exponent}) corruption.
      }
      \label{fig:gaussian-rejection-mean}
    \end{figure}

\begin{figure}[htbp]
      \centering
      \includegraphics[width=0.45\textwidth]{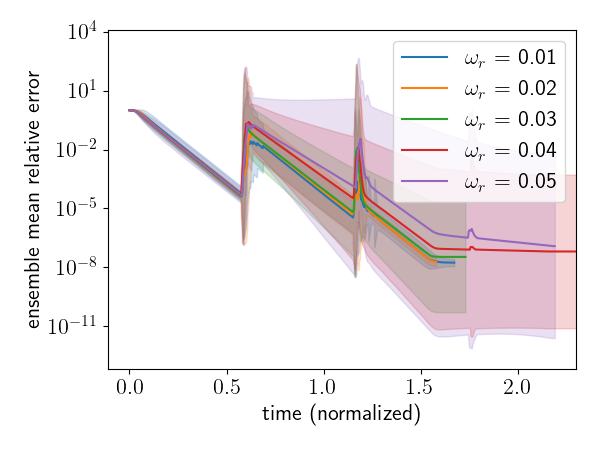}
      \includegraphics[width=0.45\textwidth]{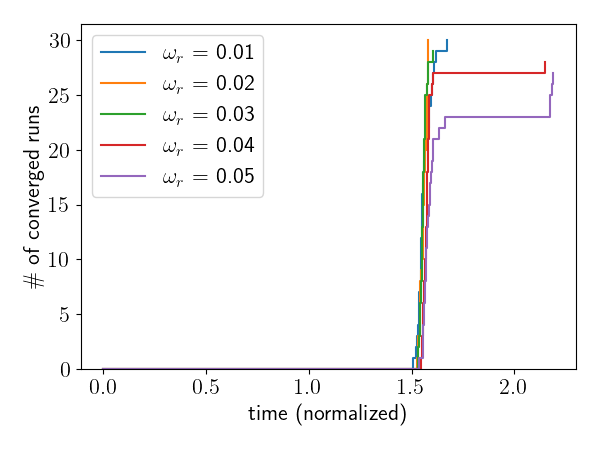}
      \caption{
        Ensemble convergence of ASJ-R with malevolent corruption with time-to-failure $\omega_f=2$s, various recovery times $\omega_r$, and offset magnitude $\delta=0.2$.
        Convergence is achieved for all runs with $\omega_r=0.01$s and $\omega_r=0.02$, and almost all runs with larger $\omega_r$, resulting in times to solution around $1.5$x longer than without corruption.
      }
      \label{fig:gaussian-rejection-recovery}
    \end{figure}

    In contrast, all of the ASJ-R runs in Figure~\ref{fig:gaussian-rejection-mean} converge for recovery time $\omega_r=0.02$s and various offset magnitudes $\delta$.
    As with the ASJ results in Figure~\ref{fig:gaussian-original}, the solution error does increase with the arrival of the first degraded state in agent $9$; however, the values of $\tilde{s}_i(t)$ on neighboring nodes have reached large enough values to reject at least some of the corruption and limit the jump in error.
    By the time agent $9$ enters the second degraded state, the values of $\tilde{s}_i(t)$ on neighboring nodes are such that more corruption is rejected than during the first degraded state, resulting in an even smaller jump in error than the first degraded state.
    Figure~\ref{fig:gaussian-rejection-recovery} shows similar progressive limiting of solution error increases during degraded states for various recovery times $\omega_r$ with offset magnitude $\delta=0.2$.
    Note that some of the runs with longer recovery times $\omega_r \geq 0.03$s do either take longer to converge or fail to converge.
    Both behaviors are explained by the longer recovery times leading to more corruption occurring during the first degraded state that can be small enough in magnitude to avoid rejection yet large enough to drive up the error on all agents.
    As we saw in Section~\ref{sec:natural-corruption}, corruption leads to either a stagnation period that is eventually corrected, leading to delayed convergence, or to enough deviation from theory that the rejection criterion rejects all further updates, leading to non-convergence.
    That said, the ASJ-R method restores convergence in almost all runs with the time-to-failure $\omega_f=2$s, at least for the recovery times and offset magnitudes selected

\subsection{Path-Length Rejection Considerations}

  Recall that the ASJ-R rejection criterion \eqref{eq:rejection} is developed on theory that uses the exact shortest path length $s_i(t)$, which is typically not available to the agents and therefore replaced by an approximation $\tilde{s}_i(t)$.
  We saw in Section~\ref{sec:rejection-results} that whether $\tilde{s}_i(t)$ in \eqref{eq:rejection} is sufficiently large to reject significant corruption at a given time $t$ has a profound impact on the convergence of ASJ-R, ranging from a temporary stagnation period that results in a longer time-to-solution to persistent stagnation that prevents convergence all together.
  While a more rigorous study is warranted for future work, the values of $\tilde{s}_i(t)$ as defined in Algorithm~\ref{alg:rejection} are found to consistently underestimate the values of $s_i(t)$ for runs that were anecdotally selected.
  As such, one might both significantly reduce the ASJ-R time-to-solution and increase the likelihood of convergence in the presence of corruption with a more accurate approximate shortest path length $\tilde{s}_i(t)$ that reduces or eliminates the stagnation issues in Section~\ref{sec:rejection-results}.

\begin{figure}[htbp]
    \centering
    \includegraphics[width=0.45\textwidth]{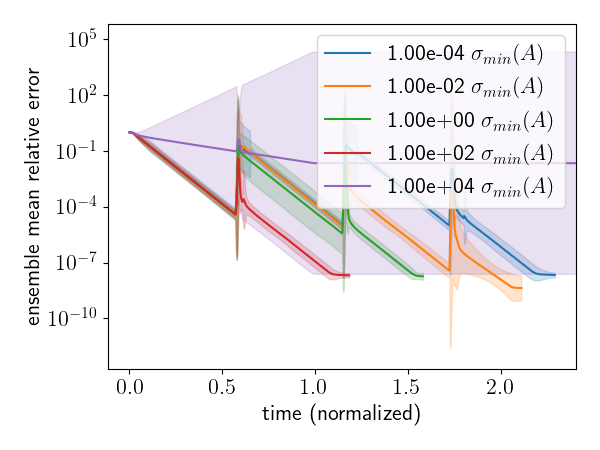}
    \includegraphics[width=0.45\textwidth]{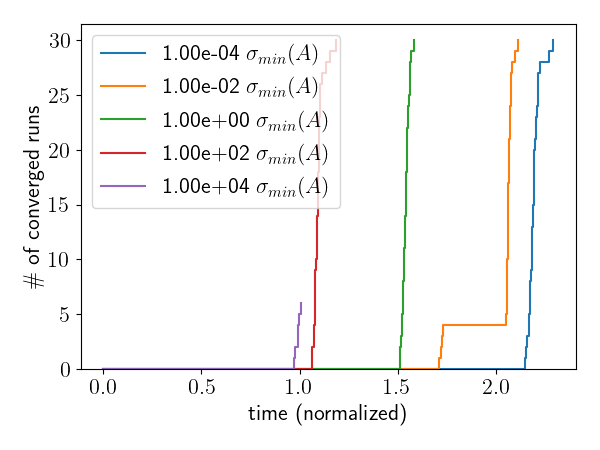}
    \caption{
      Ensemble convergence for ASJ-R with malevolent data corruption using various scaled values of $\sigma_\text{min}(A)$ in the rejection criterion \eqref{eq:rejection}.
      Convergence is attained in almost all runs for all scaling factors except $10^{4}$ scaling, with times to solution values improving for larger scaling factors.
    }
    \label{fig:singular-vals-minA}
 \end{figure}

  Another consideration for the practical use of ASJ-R is the dependence of the rejection criterion \eqref{eq:rejection} on singular values.
  For the system sizes considered in Section~\ref{sec:rejection-results}, the values of $\sigma_\text{min}(A) \approx 0.0447$ and $\sigma_\text{max}(M) \approx 0.989$ are relative cheap to compute locally on each agent; however, one might want to apply ASJ-R to large systems or to systems where agents do not have access to all rows of $A$.
  As such, the malevolent corruption study with time-to-failure $\omega_f=2$s, recovery time $\omega_r=0.02$s, and offset magnitude $\delta=0.2$ is repeated for ASJ-R but with either $\sigma_\text{min}(A)$ or $\sigma_\text{max}(M)$ replaced in \eqref{eq:rejection} by approximate values.
  Figure~\ref{fig:singular-vals-minA} shows the convergence behavior of ASJ-R with $\sigma_\text{min}(A)$ replaced by the numerically computed value scaled by one of $10^{-4}$, $10^{-2}$, $1$, $10^2$, or $10^4$.
 All the runs converge when the approximated singular value is smaller than the true value, where the smaller singular value results in a looser bound in the rejection criterion.
 The looser bound means $\tilde{s}_i(t)$ needs to attain larger values to reject the same data as when the true singular value is used, explaining why the smaller approximated values result in longer times to solution.
 Increasing the approximated singular values above the true value results in a tighter bound, which likely offsets some of the underestimation in the shortest path length.
 Overall this results in a shorter time to solution, almost that of ASJ-R without corruption, from the additional rejection for $10^2$ and in non-convergence too much rejection for $10^4$.
 \begin{figure}[htbp]
  \centering
  \includegraphics[width=0.45\textwidth]{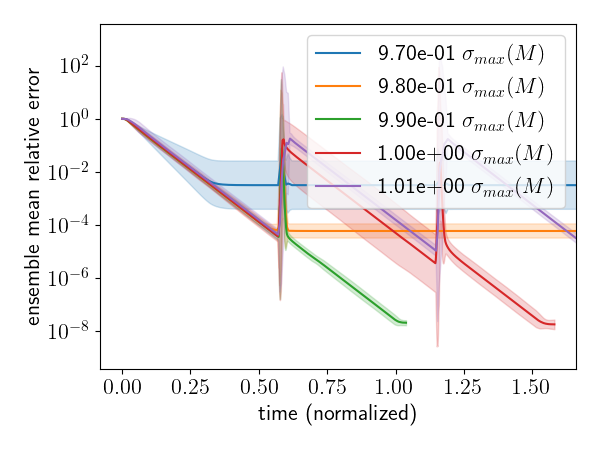}
  \includegraphics[width=0.45\textwidth]{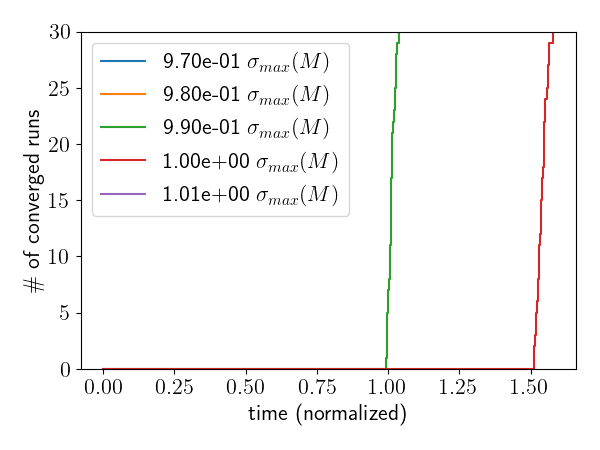}
  \caption{
    Ensemble convergence for ASJ-R with malevolent data corruption using various perturbed values of $\sigma_\text{max}(M)$ in the rejection criterion \eqref{eq:rejection}.
    Convergence is only attained for runs with scaling factors $1$ and $0.99$, with optimal times to solution for $0.99$ scaling factor.
  }
  \label{fig:singular-vals-maxM}
\end{figure}
While ASJ-R appears relatively robust to approximation of $\sigma_\text{min}(A)$, Figure~\ref{fig:singular-vals-maxM} shows the method is much less robust to $\sigma_\text{max}(M)$ being replaced by the numerically computed value scaled by one of $0.98$, $0.99$, $1$, $1.01$, or $1.02$.
When the scaling value $0.99$ is used, the resulting smaller singular value results in a tighter bound  in the rejection criterion, leading to a time to solution equal to that of ASJ-R without corruption.
All the other scaling values, however, caused the rejection criterion to be too restrictive or too passive to restore convergence.
It is worth noting that the results from Figure~\ref{fig:singular-vals-minA} and Figure~\ref{fig:singular-vals-maxM} both support the hypothesis that a better shortest path length approximation $\tilde{s}_i(t)$ will significantly improve ASJ-R performance in the presence of corruption.

\subsection{Sparsity Pattern}

\begin{figure}[htbp]
  \centering
  \includegraphics[width=0.45\textwidth]{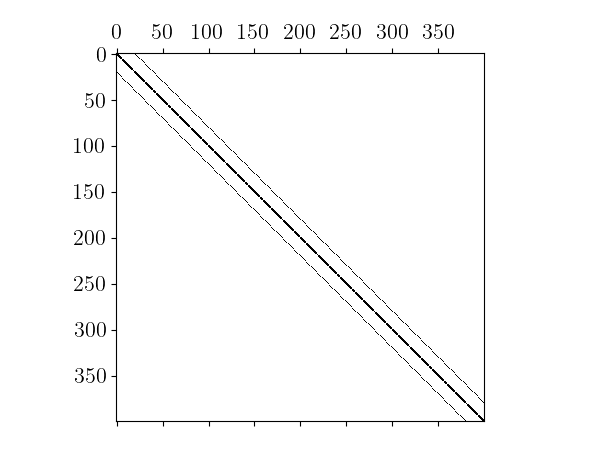}
  \includegraphics[width=0.45\textwidth]{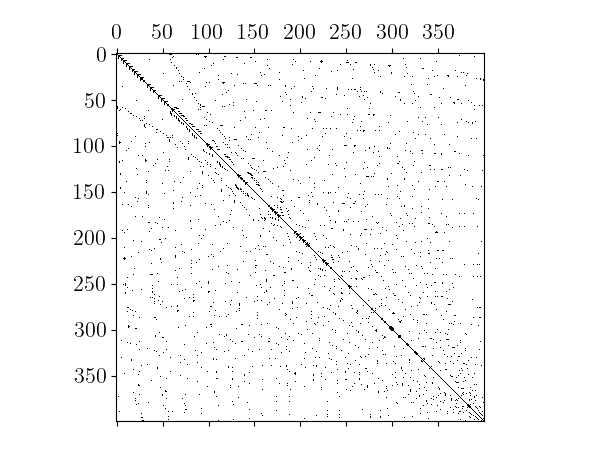}
  \caption{
    Sparsity pattern of matrix $A$ in the Poisson benchmark (left) and in the new linear system that leverages the MGG expander graph (right).
    The new linear system results in communicated data from each agent being sent to significantly more neighboring agents than the Poisson benchmark, where communicated data is sent to at most two neighboring agents.
  }
  \label{fig:sparsity-problems}
\end{figure}

The Poisson benchmark problem \eqref{eq:poisson} results in linear systems of a particular structure, i.e., the matrix $A$ in \eqref{eq:axb} has only two off-diagonal bands.
To evaluate whether the resilience to corruption seen in the prior sections extends beyond that discrete Poisson sparsity pattern, the Margulis-Gabber-Galil expander graph is leveraged from the \emph{NetworkX} software library \cite{hagberg_exploring_2008} (see \url{https://networkx.org}).
Specifically, the new linear system is defined by $A = I - G/8$ , where $I \in \R^{400 \times 400}$ is the identity matrix, $G \in \R^{400 \times 400}$ is the Margulis-Gabber-Galil (MGG) expander graph with degree $8$, and $\vec{b} \in \R^{400}$ is a vector of ones.
Figure~\ref{fig:sparsity-problems} shows that the sparsity pattern of such $A$ results in substantially more connectivity between agents compared to that of the Poisson benchmark.

\begin{figure}[htbp]
  \centering
  \includegraphics[width=0.45\textwidth]{rejection-length_bitflip_frequency_comparison_poisson-2d_successes}
  \includegraphics[width=0.45\textwidth]{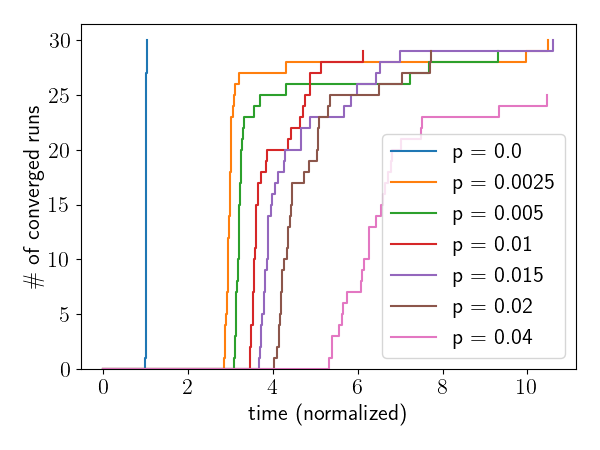}
  \caption{
    Ensemble convergence of ASJ-R for the Poisson benchmark (left) and the MGG system (right) with bit flip corruption probabilities ranging from $p=0$ to $p=0.04$, with double floating point flips in any of the $64$ bits.
    The convergence behavior is comparable, with increasing $p$ for the MGG system resulting in longer relative times to solution and a larger likelihood that a given run does not converge.
  }
  \label{fig:bitflip-probabilities-problems}
\end{figure}

\begin{figure}[htbp]
  \centering
  \includegraphics[width=0.45\textwidth]{rejection-length_gaussian_mean_comparison_poisson-2d_successes}
  \includegraphics[width=0.45\textwidth]{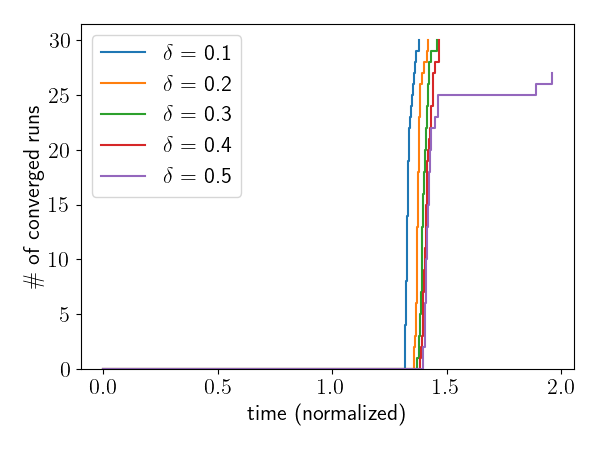} \\
  \includegraphics[width=0.45\textwidth]{rejection-length_gaussian_recovery_comparison_poisson-2d_successes}
  \includegraphics[width=0.45\textwidth]{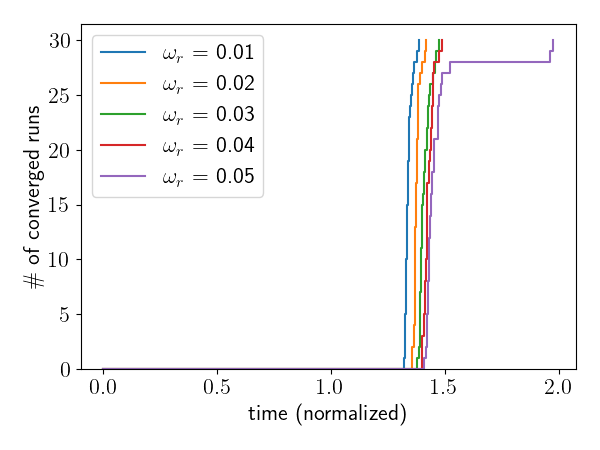}
  \caption{
    Ensemble convergence of ASJ-R in the presence of malevolent corruption for the Poisson benchmark (left column) and the MGG system (right column) with recovery time $\omega_r=0.02$s and various offset magnitudes $\delta$ (top row) and various recovery times $\omega_r$ and offset magnitude $\delta=0.2$ (bottom row).
    The convergence behavior is comparable for the two linear systems.
  }
  \label{fig:gaussian-rejection-problems}
\end{figure}

A convergence duration study is first conducted for the new linear system, as was done in Section \ref{sec:convergence_duration}, to confirm that $1$s should still be used.
The resulting average time to convergence ($\approx 2.9$s) is used to normalize the time in the new linear system.
Next, bit flips in any of the $64$ bits of the floating point double are introduced to communicated data with the same probabilities as those used for the Poisson benchmark.
Figure~\ref{fig:bitflip-probabilities-problems} shows that the ASJ-R method retains similar resilience to the bit flip corruption for the new linear system as shown for the Poisson benchmark.
That said, the increased connectivity in the new linear system does seem to result in longer relative times to solution, and a larger likelihood that a given ASJ-R run does not converge, as the probability $p$ of bit flips is increased.
This behavior is likely due to how a corrupted broadcasted solution approximation will be received by more neighbors due to the increased connectivity.
Malevolent manipulation of stored data on agent $i=9$ is also explored using the same time-to-failure $\omega_f$, recovery times $\omega_r$, and offset magnitudes $\delta$ as used in the Poisson benchmark.
Figure~\ref{fig:gaussian-rejection-problems} shows that the ASJ-R method retains resilience to the malevolent corruption for the new linear system, with the relative times to solution being slightly faster than those for the Poisson benchmark.
Whereas the increased connectivity is a disadvantage for bit flip corruption, the connectivity here provides a faster mechanism to correct the corruption both on agent $i=9$ and on any other agents that failed to reject the corrupted updates.

\section{Conclusions}
\label{sec:conclusions}
We introduced a fault-tolerant asynchronous Jacobi (ASJ) variant that leverages ASJ convergence theory by Hook and Dingle \cite{hook_performance_2018} to provide resilience to data corruption.
The resulting variant ASJ method (ASJ-R) rejects solution approximations from neighbor nodes if the distance between two successive approximations violates an analytic bound.
Because the analytic bound requires the shortest path length, the ASJ-R method includes a shortest path length approximation.
Following the work of Anzt et al. \cite{anzt_2015}, we studied the resilience of ASJ and ASJ-R to corruption in communicated data due to bit flips in various parts of the \textsc{ieee} 754 floating point representation.
While we observed that both ASJ and ASJ-R reliably converge when the corruption is very small relative to the convergence tolerance, the ASJ-R method generally retains the ability to converge when bit flips occur in locations that cause larger magnitude corruption.
The convergence of the ASJ-R exhibits a stagnation period, a degraded convergence rate, or both depending on the properties of the corruption, resulting in times to solution around $1.5$x to $6$x longer than without corruption.
By individually studying particular locations for bit flips, we are able to explain both convergence behaviors, as well as the lack of convergence, by whether the shortest path length increases at a sufficient rate to reject the errors that would otherwise delay or prevent convergence.
Stagnation periods occur while the shortest path length increases to a value needing to start rejecting the corruption.
A degraded convergence rate occurs when the corruption magnitude is small enough to avoid the rejection criterion (until the shortest path length is large enough) but large enough to slow, but not stall, the convergence.
Non-convergence occurs when the corruption that avoids the rejection criterion is enough for the solution approximations to deviate substantially from that predicted by convergence theory, resulting in the ASJ-R method rejecting the large updates that would otherwise drive the solution approximations back towards the exact solution.

We also studied the resilience of ASJ and ASJ-R to the corruption of stored data, where the stored values are perturbed by a uniformly distributed amount for periodic windows of time.
Whereas ASJ failed to converge in all the scenarios tested, ASJ-R reliably restored convergence in a large majority of those scenarios.
As with the bit flip corruption, we observed that the convergence of ASJ-R depends on whether the shortest path length approximation increased at a rate sufficient to prevent the stored data corruption from driving the solution approximations too far from that predicted by the convergence  theory upon which the rejection criterion is derived.
Given the importance of the shortest path approximation and the ASJ-R rejection criterion bound that it appears in, we studied the sensitivity of the method to approximations in the values used for the two singular values in the bound.
We found the results to be more sensitive to the maximum singular value of the iteration matrix and less sensitive to the minimum singular value of the linear system matrix, which is promising as the latter is typically more difficult to obtain.
Skewing of either singular values in the direction that tightened the rejection criterion bound was found to more likely maintain, or even improve, convergence behavior than skewing that loosened the bound.
We also verified that the ASJ-R performance extends to a second linear system constructed using the Margulis-Gabber-Galil expander graph for a more dense sparsity pattern.

While this work focused on a solving a linear system with a Jacobi method in an HPC environment with an empirically determined convergence duration, the key observations should have applicability to other environments and solvers.
Edge computing environments will very likely have greater communication latency than the HPC environment used here; however, the objective for the rejection criterion remains to reject the corruption that would cause the solution approximation to deviate too far for the rejection criterion to be useful.
While the greater communication latency will result in the shortest path length increasing at a slower rate, the latency will also slow the rate at which the corruption causes the solution approximations to deviate.
Thus, one might expect the relative times to solution observed in this work to have some relevance in edge environments.
The particular convergence duration empirically determined in this work, however, will likely not have relevance in edge computing environments.
Instead, one might empirically determine the appropriate convergence duration for a given environment in the same manner as it was determined here.
One might also \textit{a priori} leverage information about a given environment to determine a suitable convergence duration consisting of a safety factor multiplying the combination of (i) the time for a node to produce an update after reporting local convergence, (ii) the time to communicate that update to a neighbor, (iii) the time for the neighbor to produce an update and determine it is no longer locally converged, and (iv) the for the neighbor to report to the other nodes.
Finally, the particular rejection criterion derived here is indeed unique to the convergence theory for the Jacobi method.
That said, the approach of systematically evaluating a dynamic rejection bound by introducing  corruption of various magnitudes to broadcasted data can be leveraged by new asynchronous solvers as they are developed.

\bibliographystyle{siamplain}
\bibliography{references}

\end{document}